\documentclass[12pt,preprint]{aastex}

\usepackage{graphicx}
\usepackage{comment}
\usepackage{amsmath}
\usepackage{epstopdf}

\def\beq{\begin{equation}}
\def\eeq{\end{equation}}
\def\beqar{\begin{eqnarray}}
\def\eeqar{\end{eqnarray}}

\def\avg#1{\langle #1 \rangle}

\def\csfr{\dot{\rho}_\star}

\def\la{\mathrel{\mathpalette\fun <}}
\def\ga{\mathrel{\mathpalette\fun >}}
\def\fun#1#2{\lower3.6pt\vbox{\baselineskip0pt\lineskip.9pt
  \ialign{$\mathsurround=0pt#1\hfil##\hfil$\crcr#2\crcr\sim\crcr}}}

\def\cria{{\cal R}_{\rm Ia}}
\def\crias{{\cal R}_{\rm Ia, S}}
\def\criap{{\cal R}_{\rm Ia, Q}}

\def\ria{R_{\rm Ia}}
\def\rias{R_{\rm Ia, S}}
\def\riap{R_{\rm Ia, Q}}
\def\rcc{R_{\rm CC}}

\def\gammaLa{L_{\rm H {\alpha}}}
\def\Laz{L_{\rm H{\alpha,z}}}
\def\Lazold{L_{\rm H{\alpha, z+\Delta z}}}

\begin{document}

\title{The Diffuse Gamma-ray Background from Type Ia Supernovae}

\author{Amy Lien\altaffilmark{1} and Brian D. Fields\altaffilmark{2}}

\affil{Department of Astronomy, University of Illinois, Urbana, IL 61801, USA}

\altaffiltext{1}{NASA Postdoctoral Program Fellow, Goddard Space Flight Center, Greenbelt, MD 20771, USA}
\altaffiltext{2}{Department of Physics, University of Illinois, Urbana, IL 61801, USA}

\begin{abstract}

The origin of the diffuse extragalactic gamma-ray background
(EGB)
has been intensively studied but remains unsettled.
Current popular source candidates include unresolved 
star-forming galaxies, starburst galaxies, and blazars.
In this paper we calculate the EGB contribution
from the interactions of cosmic rays accelerated by 
Type Ia supernovae (SNe), extending earlier work which only
included core-collapse SNe.
We consider Type Ia events not only
in star-forming galaxies, but also in quiescent galaxies
that lack star formation.
In the case of star-forming galaxies,
consistently including Type Ia events 
makes little change to the star-forming EGB prediction, 
so long as both SN types have the same 
cosmic-ray acceleration efficiencies in star-forming galaxies.
Thus, our updated EGB estimate continues to
show that star-forming galaxies can represent a substantial portion of the
signal measured by {\em Fermi}.
In the case of quiescent galaxies,
conversely,  we find a wide range of possibilities 
for the EGB contribution. 
The dominant uncertainty we investigated comes
from the mass in hot gas in these objects, 
which provides targets for cosmic rays;
total gas masses are as yet poorly known,
particularly at larger radii.  
Additionally, 
the EGB estimation is very sensitive to
the cosmic-ray acceleration 
efficiency and confinement, especially in quiescent galaxies.
In the most optimistic allowed scenarios,
quiescent galaxies can be an important source of the 
EGB.
In this case, star-forming galaxies and quiescent galaxies together 
will dominate the EGB and leave little room for other contributions.
If other sources, such as  blazars,
are found to have important contributions to the EGB,
then either the gas mass or cosmic-ray content of
quiescent galaxies must 
be significantly lower than in their star-forming counterparts.
In any case, improved {\em Fermi} EGB measurements will provide 
important constraints on hot gas
and cosmic rays in quiescent galaxies. 

\end{abstract}

\section{Introduction}

The first observation of the diffuse extragalactic gamma-ray
background (EGB) was reported by
the SAS-2 satellite \citep{Fichtel77, Fichtel78}.
Recently, the {\it Fermi Gamma-ray Space Telescope} 
updated the EGB determination
from the {\em Energetic Gamma-ray Experiment Telescope} \citep{Sreekumar98}
and provided the most reliable EGB observations so far
\citep{Abdo09_fermiEGB}.
Measurements of EGB
are model dependent in that
they require subtraction of the
large foreground emission from our Galaxy \citep[e.g.,][]{Hunter97}.
The accuracy of the EGB measurement thus
greatly depends on our understanding of the Galactic emission.
Despite the difficulty in its observation, 
the EGB encodes important information 
about the highest-energy environments in the cosmos.

The EGB arises from the combination of all the 
unresolved extragalactic gamma-ray sources \citep[e.g.,][]{Dermer07, Stecker10}.
``Guaranteed'' EGB components arise from
unresolved counterparts of known extragalactic 
populations, namely, 
blazars
\citep[those active galactic nuclei that have their relativistic jets pointing at us,
e.g.,][]{Padovani93, Stecker93, Mukherjee99, Pavlidou08, Dermer07AGN,Venters10,Venters11,Inoue09},
as well as normal star-forming galaxies and starburst galaxies 
\citep[e.g.,][]{Pavlidou01, Pavlidou02,
Prodanovic06,Thompson07, Stecker07,Fields10,Makiya11}.
Additional EGB contributions might arise from
more exotic sources, such as
dark matter annihilation \citep{Silk84, Rudaz91},  
annihilations at the boundaries of
cosmic matter and antimatter domains \citep{Stecker71_matter},
massive black holes at redshifts of $z \sim 100$ \citep{Gnedin92},
and primordial black hole evaporation \citep{Page76}.

In this paper we will focus on the EGB contribution from
both star-forming galaxies and quiescent galaxies.
Quiescent galaxies refer to galaxies with 
little or no active star formation,
and these objects have not been included in EGB estimations. 
In terms of galaxy types, quiescent galaxies usually include all elliptical galaxies and 
some S0 galaxies. However, the important factor for the EGB estimation
is not the galaxy type but the amount of star formation.
Therefore we will separately consider
star-forming and quiescent galaxies,
and assume no star formation in quiescent galaxies.
We will not consider the EGB contribution from starburst galaxies
in this paper, due to the larger uncertainty in the cosmic-ray propagation in
such galaxies \citep[e.g.,][]{Thompson07,Lacki10}.
Also, recent work suggests that starburst galaxies only have
small contribution to the EGB \citep{Stecker10}.
We follow the criterion adopted in \citet{Fields10} 
to distinguish star-forming and starburst galaxies.
The EGB energy range we consider in this paper is from $\sim 30$ MeV to $\sim 30$ GeV,
which is the energy range covered by {\it Fermi} data
and includes the regime in which star-forming galaxies may 
contribute substantially to the EGB.

Our focus here is on the EGB contribution arising
from hadronic cosmic-ray interactions with the interstellar medium (ISM)
of their host galaxies, specifically pion production and
decay $\rm pp \rightarrow \pi^0 \rightarrow \gamma \gamma$. 
The most favored possibility of the cosmic-ray 
production sites in galaxies is supernovae (SNe).
{\em Fermi} and air \v{C}erenkov  observations
detect individual remnants of both
core-collapse
\citep[hereafter CC;][]{Abdo10_CasA,Acciari11_CasA,Weekes89} and Type Ia events
\citep{Acciari11_Tycho,Acero10}.  
The energetics of these objects are consistent with
the requirements of efficient cosmic-ray acceleration
\citep[e.g.,][]{Abdo10_CasA,Reynolds92}, and the GeV spectra of these objects are 
consistent with pionic emission and thus
hadronic acceleration \citep[e.g.,][]{Abdo09_W51C,Tanaka11,Ellison12}.
Diffuse Galactic emission is similarly consistent with pionic emission
dominating \citep{Abdo09_fermiEGB,Abdo09_CRint}.
These data thus give empirical grounding to the long-held
belief that SNe of all types are the dominant 
engines that accelerate Galactic cosmic rays 
\citep[e.g.,][]{Baade34,Ginzburg64,Ellison97}.
We are thus interested 
in SNe of both types, all of which accelerate cosmic rays
in their host galaxies.

Many groups have studied the EGB emission from
cosmic rays accelerated by
SNe in star-forming galaxies \citep[e.g.,][]{Dar95,Prodanovic06,Fields10,Stecker10,Makiya11}.
Some estimations suggest that
star-forming galaxies can be the dominant source of the EGB \citep{Fields10},
while other groups predict that a major contribution of the EGB comes from blazars \citep{Stecker10,Makiya11,Inoue09}.
However, there exist large uncertainties from the source inputs.
Most of the analyses regarding star-forming galaxies focused on
the EGB contribution from cosmic rays accelerated by CC SNe
and implicitly assume that {\em only} these events accelerate cosmic rays. 
We extend the analysis of the EGB from star-forming galaxies in \citet{Fields10}
to include Type Ia SNe as accelerators in the Milky Way and in other galaxies.

CC SNe arise in massive stars with short lifetimes, 
and thus trace ongoing star formation.  
In contrast,
Type Ia SNe result from thermonuclear runaway of white dwarfs accreting mass from their 
companion stars and hence are related to star formation with some delay time.
For this reason, observations have shown that Type Ia SNe exist in both 
star-forming galaxies and quiescent galaxies, while CC SNe are rarely seen in quiescent galaxies 
\citep{Filippenko01,Mannucci05}.
Observations have suggested that the intrinsic cosmic CC SN rate is 
about 5  times higher than the intrinsic cosmic Ia SN rate at redshift $z < 0.4$ \citep{Bazin09}.
Also, studies suggest that the Ia rate in a star-forming galaxy 
is much larger than that in a quiescent galaxy; this reflects
the distribution of delay times between progenitor birth and Ia explosion,
which is weighted towards short delays \citep{Mannucci05,Sullivan06},

The efficiency of cosmic-ray acceleration by SNe remains poorly understood 
but is crucial for understanding cosmic-ray acceleration physics as well as 
SN energy feedback.
Theories propose that cosmic rays are produced by diffusive shock acceleration 
in the blast waves from SN explosions \citep[e.g.,][]{Schlickeiser89, Berezhko99}.
Current studies suggest that $\sim 30\%$ of the initial 
kinetic energy from a SN needs to be transferred to
cosmic-ray acceleration if we assume that SNe 
are the dominate sources for cosmic-ray production 
and the nucleosynthesis of lithium, beryllium, and boron
in the Milky Way \citep{Fields01}.
Also, some theoretical predictions expect 
the cosmic-ray acceleration efficiency 
in quiescent galaxies is much lower than in star-forming galaxies. \citet{Dorfi96}
suggest that only $\la 1\%$ of the total explosion energy goes into cosmic-ray energy in quiescent galaxies.  
This is because a SN blast will have weaker shocks
due to the large sound speeds of the
hot, low-density ISM
in an elliptical galaxy.

Understanding the SN rate and their efficiency in producing cosmic rays is
critical for studying the EGB contributions from 
these galaxies.
Our observational understanding of cosmic SNe will increase significantly
when the next generation optical survey telescope,
the Large Synoptic Survey Telescope (LSST),
comes online during the next decade.
LSST is planning to scan the whole available sky, repeated 
every $\sim 3$ days, with unprecedented survey sensitivity 
\citep{LSST_overview}.
The project will observe $\sim 10^5$ CC SNe
per year out to redshift $z \sim 1$ \citep{lf}
and $\sim 5 \times 10^4$ Type Ia events out to redshift $z \sim 0.8$ \citep{LSSTsb_Ia}.
The cosmic SN rate in different galaxy classes can thus be measured via {\it direct counting} to high redshift with 
extremely low statistical uncertainty.

In this paper, we will first describe the general formalism 
of estimating the EGB from cosmic rays accelerated by SNe in both 
star-forming and quiescent galaxies (\S~\ref{sect:gamma_formalism}).
We will then discuss the cosmic Type Ia rate in each galaxy 
classification that will be used in our EGB analysis (\S~\ref{sect:gamma_Ia}).
The estimations of the EGB contribution from Type Ia SNe 
in star-forming and 
quiescent galaxies are presented in \S~\ref{sect:gamma_Ia_S} and \S~\ref{sect:gamma_Ia_E}, 
respectively.
Additionally, we discuss the uncertainties
in the EGB predictions in \S~\ref{sect:gamma_uncertainty}.
Finally, we summarize the results in \S~\ref{sect:gamma_conclusion}. 

\section{General Formalism}

\label{sect:gamma_formalism}

The formalism we adopt generalizes that of
\citet{Fields10} to account for both SN types.
Integration of the gamma-ray contributions from 
each unresolved extragalactic source over the line of sight to the cosmic horizon
gives the well-known express for
the EGB intensity,
\beq
\label{eq:gamma_general}
\frac{dI}{dE} = \frac{c}{4 \pi} \int {\cal L}_{\gamma}(E_{\rm em},z) \
		(1+z) \left|\frac{dt}{dz}\right| \ dz,
\eeq 
where ${\cal L}_{\gamma}(E_{\rm em},z)= dN_{\rm gamma}/dV_{\rm com} \, dt_{\rm em} dE_{\rm em}$ is the comoving luminosity density 
(or emissivity) at rest-frame energy $E_{\rm em}$, and 
$|dt/dz| = [(1+z)H(z)]^{-1} = [(1+z) H_0 \sqrt{\Omega_m(1+z)^3+\Omega_{\Lambda}}]^{-1}$ for the standard $\Lambda$CDM cosmology.
We use $\Omega_m = 0.27$, $\Omega_{\Lambda}=0.73$,
and $H_0 = 70 \ \rm km \ s^{-1} \ Mpc^{-1}$ 
from the seven-year {\it Wilkinson Microwave Anisotropy Probe} ({\it WMAP}) data \citep{wmap7}.

Because the pionic gamma-ray emission is produced from
the interaction between cosmic rays and 
the hydrogen atoms in the ISM of each galaxy,
the luminosity density is given
by the product
\beq
{\cal L}_{\gamma} = \avg{L_\gamma \ n_{\rm gal}},
\eeq
of 
the pionic gamma-ray luminosity $L_\gamma$ of an individual galaxy,
times the galaxy number density, appropriately averaged.
Our problem then divides into two parts.  First,
we must express a galaxy's gamma-ray luminosity
$L_\gamma$ in terms of
galaxy properties such as SN rate and gas content,
and relate these to galaxy observables.
Then we must construct a luminosity function $dn_{\rm gal}/dL_\gamma$
for gamma-ray-emitting galaxies.

We first turn to the pionic gamma-ray luminosity 
from an individual galaxy. 
This can be written as
\begin{align}
L_\gamma (E_{\rm em}) &= \int \Gamma_{\pi^0 \rightarrow \gamma \gamma}(E_{\rm em}) \ n_{\rm H} \ dV_{\rm ISM} \\
			&= \Gamma_{\pi^0 \rightarrow \gamma \gamma}(E_{\rm em}) \ {\cal N}_{\rm H} 
\end{align}
where
$\Gamma_{\pi^0 \rightarrow \gamma \gamma}(E_{\rm em})$ represents a spatial average 
of the gamma-ray production rate per interstellar hydrogen atom.
The total number ${\cal N}_{\rm H} = \int n_{\rm H} \ dV_{\rm ISM}$ 
of hydrogen atoms in a galaxy is
obtained by integrating the number density of hydrogen atom $n_{\rm H}$ over the ISM volume.
${\cal N}_{\rm H}$ is proportional to the total gas mass $M_{\rm gas}$ in a galaxy 
and can therefore be expressed as ${\cal N}_{\rm H} = X_{\rm H} \ M_{\rm gas}/m_p$,
where $X_{\rm H}$ is the mass fraction of hydrogen atoms and $m_p$ is the proton mass. 

We take SNe (of both types)
to be the engines of cosmic-ray acceleration;
this implies that
the cosmic-ray 
flux scales as $\Phi_{\rm cr} \propto \Lambda_{\rm esc} R_{\rm SN,eff}$.
Here 
$R_{\rm SN,eff}$
is an effective SN rate weighted by the cosmic-ray acceleration
efficiency $\epsilon$, discussed below.
$\Lambda_{\rm esc}$ is the escape path length,
which quantifies the cosmic-ray confinement in a galaxy.
In this paper we assume $\Lambda_{\rm esc}$ 
to be universal and constant, which
leads to a universal galactic cosmic-ray spectrum
that is the same as that of the Milky Way.
Thus the pionic gamma-ray production rate per hydrogen atom
in a galaxy should scale as
$\Gamma_{\pi^0 \rightarrow \gamma \gamma}(E_{\rm em}) \propto \Phi_{\rm cr} \propto \ \Lambda_{\rm esc} \ R_{\rm SN,eff}$.
We normalize
the cosmic-ray spectrum to a known galaxy, which would be the Milky Way in our case, finding
\beq
\frac{\Gamma_{\pi^0 \rightarrow \gamma \gamma}(E_{\rm em})}{\Gamma^{\rm MW}_{\pi^0 \rightarrow \gamma \gamma}(E_{\rm em})}
 =\frac{\Phi_{\rm cr}}{\Phi^{\rm MW}_{\rm cr}}         
  =\frac{R_{\rm SN,eff}}{R^{\rm MW}_{\rm SN,eff}}.
\eeq
The pionic gamma-ray luminosity of a particular galaxy is thus
\beq
\label{eq:gamma_tot}
{L_\gamma (E_{\rm em})} = \Gamma^{\rm MW}_{\pi^0 \rightarrow \gamma \gamma}(E_{\rm em}) \ \frac{R_{\rm SN,eff}}{R^{\rm MW}_{\rm SN, eff}} \ X_{\rm H} \ 
  \frac{M_{\rm gas}}{m_p}.
\eeq 
Because of their short lifespans, the rate of CC SNe (and short-delay
Ia events) traces that
of star formation:   $R_{\rm SN} \propto \psi$.
Thus we expect a star-forming galaxy's gamma-ray luminosity
to scale as $L_\gamma \propto M_{\rm gas} R_{\rm SN} \propto M_{\rm gas} \psi$.
The new class of {\em Fermi}-detected star-forming galaxies
is consistent with this trend \citep{Lenain11}.

This pionic gamma-ray spectrum always has a peak at $E_{\rm em} = m_{\pi^0}/2$,
at which the two gamma-ray photons inherit the rest-mass energy
of the decayed $\pi^0$ \citep{Stecker71}.
At large energy, the spectrum shows the same asymptotic index as that of the cosmic-ray spectrum, 
which we take to be 2.75.

Both CC and Type Ia events should produce cosmic rays
and hence pionic gamma rays.
Therefore, the effective SN rate $R_{\rm SN,eff}$ in Eq.~(\ref{eq:gamma_tot}) 
is a combination of the effective Type Ia rate ${R_{\rm Ia,eff}} \equiv \epsilon_{\rm Ia} \ \ria$ 
and the effective CC SN rate $R_{\rm CC,eff} \equiv \epsilon_{\rm CC} \ \rcc$, 
where $\epsilon_{\rm Ia}$ and $\epsilon_{\rm CC}$ are 
the cosmic-ray production efficiencies of
Type Ia and CC SNe, respectively.
There exist different definitions 
of the cosmic-ray acceleration efficiency in current literature. 
For example, some studies present the efficiency as 
the fraction of the total cosmic-ray production energy
out of the total kinetic energy output from a SN \citep[e.g.,][]{Dorfi96,Fields01,Helder10},
while other studies define the parameter as 
the percentage of the energy flux that becomes relativistic particles
after crossing the shock \citep[e.g.,][]{Ellison07}. 
Most of these definitions describe the fraction of the SN explosion energy
transferred to cosmic rays.
Here, we define the cosmic-ray acceleration efficiency $\epsilon$ to be the 
ratio of the SN baryonic explosion energy $E_{\rm SN}$ to
the resulting cosmic-ray energy $E_{\rm cr}$,
i.e., $\epsilon = E_{\rm cr}/ E_{\rm SN}$.
Therefore, if we assume that all SNe have the same explosion energy
and the produced cosmic rays have the same energy spectrum,
the cosmic-ray acceleration efficiency will be proportional to
the total cosmic-ray production in a galaxy over the SN rate in that galaxy, i.e.,
$\epsilon \propto \Phi_{\rm cr} / (\Lambda_{\rm esc} \ R_{\rm SN})$.
For the Milky Way, then, we have 
$R_{\rm SN,eff}^{\rm MW} = \epsilon_{\rm Ia, MW} \ R^{\rm MW}_{\rm Ia} + \epsilon_{\rm CC,MW} \ R^{\rm MW}_{\rm CC}$.

Since we normalized our prediction to the gamma-ray production in the Milky Way (Eq.~\ref{eq:gamma_tot}),
the important factor in the calculation is not the absolute value of $\epsilon$, but 
the difference between the acceleration efficiency $\epsilon$ in different 
SN types (Ia and CC)
and galaxy classes (quiescent and star-forming). 
Specifically, we will need to specify the
ratios $\epsilon_{\rm Ia} / \epsilon_{\rm CC}$ 
and $\epsilon_{\rm Q} / \epsilon_{\rm S}$.
Unfortunately, these two fractions are poorly known. 
Thus for our fiducial numerical results, we will take $\epsilon_{\rm Ia} / \epsilon_{\rm CC} = 1$
and $\epsilon_{\rm Q} / \epsilon_{\rm S} = 1$.
Furthermore, we are unaware of any evidence for a substantial difference between
the cosmic-ray acceleration efficiencies between Ia and CC SNe,
and thus we will hereafter drop the Ia and CC notation in the acceleration efficiency to simplify the discussion.
However, we will retain the notations of the acceleration efficiencies 
for different galaxy type $\epsilon_{\rm Q}$ and $ \epsilon_{\rm S}$
in our formalism as an explicit reminder that 
the efficiencies are likely to depend on galaxy environment, 
as expected by some theoretical analyses \citep{Dorfi96, Hein08}.
Further possibilities of choosing different cosmic-ray acceleration efficiencies
will be discussed in \S~\ref{sect:gamma_uncertainty_Q}.

Star-forming galaxies contain both Type Ia and CC SNe. 
Their pionic gamma-ray luminosity density 
${\cal L}_{\gamma, \rm S}$ can 
be calculated by averaging over the galaxy density $n_{\rm galaxy}$,
\beq
\label{eq:gamma_L_tot_S}
{\cal L}_{\gamma, \rm S} 
  = \Gamma^{\rm MW}_{\pi^0 \rightarrow \gamma \gamma}(E_{\rm em})
 \ \frac{X_{\rm H}}{m_p} \ 
\frac{ \avg{M_{\rm gas} \ \epsilon_{\rm S} \rias \ n_{\rm galaxy}}+\avg{M_{\rm gas} \ \epsilon_{\rm S} \rcc \ n_{\rm galaxy}} }
{\epsilon_{\rm MW} \ R^{\rm MW}_{\rm Ia} + \epsilon_{\rm MW} \ R^{\rm MW}_{\rm CC}} .
\eeq
In quiescent galaxies, there are almost no star formation. 
We will assume the star-formation rate (and thus the CC SN rate) 
to be zero in a quiescent galaxy.
However, Type Ia SNe do exist in quiescent galaxies because 
these events occur some time after the star formation.
Therefore the pionic gamma-ray luminosity density in quiescent galaxies 
${\cal L}_{\gamma, \rm Q}$ only comes from 
Type Ia events, 
\beq
\label{eq:gamma_L_tot_P}
{\cal L}_{\gamma, \rm Q} = \frac{\Gamma^{\rm Q0}_{\pi^0 \rightarrow \gamma \gamma}(E_{\rm em})}
				{\epsilon_{\rm Q0} \ R^{\rm Q0}_{\rm Ia}} \ \frac{X_{\rm H}}{m_p} \
                        \avg{M_{\rm gas} \ \epsilon_{\rm Q} \ \riap \ n_{\rm galaxy}}.
\eeq
$\Gamma^{\rm Q0}_{\pi^0 \rightarrow \gamma \gamma}(E_{\rm em})$ and $R^{\rm Q0}_{\rm SN}$ are the 
gamma-ray production rate and Type Ia event rate in a standard quiescent galaxy Q0 for normalization.
However, since no gamma-ray emission from a (jet-less) quiescent galaxy has ever been measured,
we will still adopt the values of the Milky Way and estimate the gamma-ray luminosity density for quiescent galaxies as 
\beq
\label{eq:gamma_L_tot_P_MW}
{\cal L}_{\gamma, \rm Q} = 
\Gamma^{\rm MW}_{\pi^0 \rightarrow \gamma \gamma}(E_{\rm em})
				 \ \frac{X_{\rm H}}{m_p} \
\frac{ \avg{M_{\rm gas} \ \epsilon_{\rm Q} \ \riap \ n_{\rm galaxy}} }
				{\epsilon_{\rm MW} \ R^{\rm MW}_{\rm Ia} + \epsilon_{\rm MW} \ R^{\rm MW}_{\rm CC}} .
\eeq
Note that since the gamma-ray production from the Milky Way comes from both Type Ia and CC SNe,
$\Gamma^{\rm MW}_{\pi^0 \rightarrow \gamma \gamma}(E_{\rm em})$ needs to be normalized to
the total SN rate in the Milky Way instead of just the Type Ia rate.

The total pionic gamma-ray luminosity density will be a combination of emissions from 
both star-forming and quiescent galaxies, that is, ${\cal L}_{\gamma, \rm tot} = {\cal L}_{\gamma, \rm S} + {\cal L}_{\gamma, \rm Q}$.
The EGB contribution from cosmic rays accelerated by CC SNe has been carefully examined in \citet{Fields10} and by other groups \citep[e.g.,][]{Stecker10,Makiya11}.
Here we will focus on the EGB contributions related to Type Ia events.
In our calculation, we do not include the intergalactic EGB absorption $\ga 30$ GeV \citep{Salamon98}.

\section{The Cosmic Type Ia Supernova Rate in Star-forming and Quiescent Galaxies}
\label{sect:gamma_Ia}

Type Ia SNe do not all trace ongoing star formation, because these events have different origins from CC SNe.
The prevailing scenarios for Type Ia SN origin include
merging of two white dwarfs \citep[double degenerate,][]{Webbink84},
or a white dwarf accreting from
mass-overflow of its supergiant companion \citep[single degenerate,][]{Nomoto84,Iben84}.
Both of these scenarios involve white dwarfs merging in a binary system, and thus
Type Ia SNe are delayed from the formation of the progenitor stars. 
For this reason, Type Ia SNe are found in all galaxies, including the quiescent galaxies 
where there is no longer star-forming activity.
A complete account of the Type Ia SN contribution 
to the EGB must therefore include contributions 
from events in star-forming and quiescent galaxies.

\begin{figure}[!h]
\begin{center}
\includegraphics[width=0.85\textwidth]{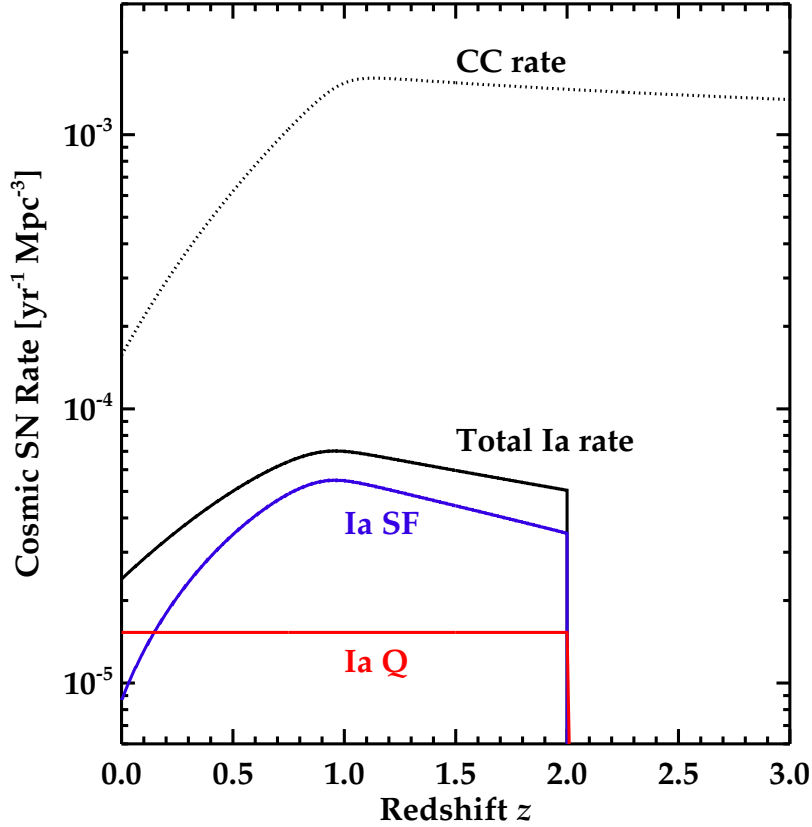}
\end{center}
\caption{
Adopted cosmic SN rate. The solid black curve plots the total
cosmic Type Ia rate; blue curve plots the
cosmic Ia rate in star-forming galaxies;
red curve plots the cosmic Ia rate in quiescent galaxies.
The cosmic CC SN rate is plotted as dotted-black curve for comparison.
}
\label{fig:gamma_Ia}
\end{figure}

There are many studies of the comoving cosmic Ia rate density 
$\cria = dN_{\rm Ia} / (dV_{\rm com} \ dt)$ as a function of redshift,
most of which focus on the distribution of delay times
\cite[e.g.,][]{Scannapieco05,Sullivan06,Kuznetsova08,Dilday10,Horiuchi10,Graur11}.
Our adopted total (all-galaxy) cosmic Type Ia rate is based on current observational data.
To find this we use the best-fit redshift and delay-time distribution
of \citet{Horiuchi10}, integrated over all delay times 
(black curve in Fig.~\ref{fig:gamma_Ia}).
Although the number of observed cosmic Ia events is rapidly
increasing, the data remain sparse beyond $z \sim 1$, where
the rates are thus poorly constrained.

The cosmic Ia rate in different galaxy classes (i.e., star-forming and quiescent galaxies)
as a function of redshift is still poorly understood.
In our calculation, we adopt a constant value for the cosmic Ia rate in quiescent galaxies out to redshift $z \sim 2$ 
(red curve in Fig.~\ref{fig:gamma_Ia}).
We use a normalization based on observations of Ia rate per stellar mass in quiescent galaxies provided by \citet{Sullivan06} 
and a non-evolving stellar-mass function from \citet{Pannella09} (see detailed discussion in \S~\ref{sect:gamma_Ia_E}).
Recently, Type Ia SNe have been observed
in galaxy clusters.  
The observed Ia rates show little redshift evolution within $z \la 1$
\citep{Gal-Yam02, Sharon07, Graham08, Mannucci08, Dilday10_Iacluster, Sharon10}. 
Since galaxy clusters are mostly composed of quiescent galaxies, 
these results are consistent with our assumption of a constant Ia rate in quiescent galaxies.
Moreover, these measurements show that Ia rate in clusters is $\sim 10^{-13} \ {\rm yr^{-1}} \ M^{-1}_{\odot}$, which is very similar to 
the value we adopted from \citet{Sullivan06}
\footnote{We are greatly thankful for the anonymous referee for pointing out the measurements of Ia rate in galaxy clusters.}.
However, while we expect a roughly constant cosmic Ia rate in quiescent galaxies for moderate redshift, this trend 
must fail at some redshift. To be conservative, we therefore placed an artificial cutoff 
of the cosmic Type Ia rate at $z = 2$, beyond which uncertainties in
the cosmic Type Ia rate observations remain substantial. 

The cosmic Ia rate in star-forming galaxies (blue curve in Fig.~\ref{fig:gamma_Ia}) can thus be obtained 
by subtracting the cosmic Type Ia rate in quiescent galaxies (red curve in Fig.~\ref{fig:gamma_Ia}) from the total cosmic Ia rate (black curve in Fig.~\ref{fig:gamma_Ia}).

Figure~\ref{fig:gamma_Ia} shows the adopted cosmic Ia SN rate as a function of redshift.
Although the uncertainty in the rate increases significantly at higher redshift,
most of the EGB from Type Ia SNe, like that from CC SNe, arises from events at lower redshift ($z \la 1$)
\citep{Ando09_EGB}.
For example, in our calculation $\sim 50\%$ ($\sim 70\%$) of the EGB flux comes from sources within 
$z \le 1$ ($z \le 1.3$). 
Therefore the choice of the Type Ia rate at $z \ga 1$ only has a small effect 
on the final estimation of the EGB.
The solid black curve plots the total cosmic Ia SN rate in both star-forming 
and quiescent galaxies.
The red curve shows the cosmic Ia SN rate in only
quiescent galaxies. 
The blue curve represents the cosmic Ia SN rate in only
star-forming galaxies.

Figure~\ref{fig:gamma_Ia} also shows the cosmic CC SN rate (dotted black curve),
which is higher than the Ia rate by a factor $\sim 5$ at $z \sim 0$
and increase to a factor of $\sim 10$ at $z \sim 1$.
This immediately suggests that we should expect CC events to
dominate the star-forming EGB signal, with the Ia contribution at a $\lesssim 20\%$ level.
We will see that this is roughly the case for the Ia contribution
from star-forming galaxies, but for Ia events in quiescent galaxies
the situation is much more uncertain.

\section{The Extragalactic Gamma-ray Background from Type Ia Supernovae in Star-forming Galaxies}

\label{sect:gamma_Ia_S}

As described in \S~\ref{sect:gamma_formalism}, 
the EGB luminosity density ${\cal L}^{\rm Ia}_{\gamma, \rm S}$
is dominated by two physics inputs: the SN rate in a galaxy, which 
is associated with the amount of cosmic rays, and the total gas mass of that galaxy,
which accounts for the total hydrogen targets that interact with the cosmic rays.
To reflect these two physics inputs, we
follow the approach adopted in \citet{Fields10} and rewrite the EGB contribution from Type Ia events (the first term in Eq.~\ref{eq:gamma_L_tot_S}) as below,
\beq
\label{eq:gamma_Ia}
{\cal L}^{\rm Ia}_{\gamma, \rm S} =  \frac{\epsilon_{\rm S} \ \crias}
					{\epsilon_{\rm MW} \ R^{\rm MW}_{\rm Ia} + \epsilon_{\rm MW} \ R^{\rm MW}_{\rm CC}} 
 \ \Gamma^{\rm MW}_{\pi^0 \rightarrow \gamma \gamma}(E_{\rm em})
					\ \frac{X_{\rm H}}{m_p} \
                                	\avg{M_{\rm gas,S}} ,
\eeq
where 
\begin{align}
\avg{M_{\rm gas,S}} &\equiv \frac{\avg{M_{\rm gas,S} \ \ \rias \ n_{\rm galaxy,S}}}{\avg{\rias \ n_{\rm galaxy,S}}} \\
		&= \frac{\int d\Laz \ M_{\rm gas,S}(\gammaLa, z) \ \rias(\gammaLa,z) \ \frac{dn}{d\Laz}}{\int d\Laz \ \rias(\gammaLa,z) \ \frac{dn}{d\Laz}},
\end{align}
and 
$\crias \equiv \avg{\rias \ n_{\rm galaxy,S}}$ 
is the cosmic Type Ia rate in star-forming galaxies,
as shown in Fig.~\ref{fig:gamma_Ia}.
In a star-forming galaxy, the galaxy gas mass $M_{\rm gas, S}$ and the galaxy Type Ia rate $\ria$
can be related to the star-formation rate in that galaxy,
which can be connected to the observable $\rm H{\alpha}$ luminosity $\Laz$ 
of the galaxy by $\psi(\gammaLa, z)/(1 \ M_{\odot} \ \ \rm yr^{-1}) = \Laz / (1.26 \times 10^{34} \ \rm W)$ 
\citep{Hopkins04}.
Therefore we express the gas mass $M_{\rm gas, S}$ and the Type Ia rate $\ria$ 
in terms of $\Laz$. The corresponding galaxy luminosity function at this wavelength
can be expressed by the Schechter function \citep{Nakamura04}.

At a specific redshift, the gas mass
in star-forming galaxies $M_{\rm gas,S}$ and the star-formation rate
can be connected by 
\beq
\label{eq:M_gas_S}
M_{\rm gas,S} =  2.8 \times 10^9 \ M_{\odot} \ (1+z)^{-\beta} \ \left(\frac{\psi}{1 \ M_{\odot} \ \rm yr^{-1}}\right)^{\omega},
\eeq
with $\beta = 0.571$ and $\omega = 0.714$,
as shown in \citet{Fields10}.
The Type Ia rate in a galaxy can be linked to the star-formation rate via some delay-time distribution $\Delta(\tau)$,
\beq
\label{eq:delay}
\ria(z) \propto \int^{t(z)}_0 \ \psi(t-\tau) \ \Delta(\tau) \ d\tau,
\eeq
where $t(z)$ is the corresponding cosmic age at redshift $z$.
The delay-time distribution  
$\Delta(\tau)$ gives the probability that a Type Ia SN explodes a time $\tau$ after the progenitor's birth.
More detailed discussion about the delay-time distribution can be found in
Appendix \ref{app:delay}.
The galaxy luminosity function at a certain redshift for star-forming galaxies in the $\rm H{\alpha}$ band
can be presented in the form of a Schechter function of
\beq
\label{eq:lumfunc_S}
\frac{dn}{d\Laz} = \frac{n_{\star,z}}{L_{\star,z}} \ 
  \left( \frac{\Laz}{L_{\star,z}} \right)^{-\alpha} \ e^{-\Laz/L_{\star,z}}
\eeq
with $\alpha = 1.43$ \citep{Nakamura04}.

Because Type Ia SNe are delayed relative to star formation,
the Type Ia rate in a galaxy depends on the
star formation history of the galaxy via eq.~(\ref{eq:delay}).
Unfortunately, this past star formation rate is not easily determined
for distant galaxies even when the H$\alpha$ luminosity 
$\Laz$ is available to give the instantaneous star formation rate.
However, we can investigate the evolution in two simplified cases: 
pure luminosity evolution and pure density evolution.
Pure luminosity evolution assumes that galaxy luminosities evolve with redshift,
while galaxy density stays unchanged, i.e., $L_{\star, z}$ in Eq.~(\ref{eq:lumfunc_S})
has redshift dependence and $n_{\star,z}$ does not.
Pure density evolution assumes that galaxy density evolves with redshift,
while galaxy luminosity remains constant,
i.e., $n_{\star, z}$ in Eq.~(\ref{eq:lumfunc_S}) depends on redshift 
and $L_{\star,z}$ does not.
The real situation should be bracketed by these possibilities.

In either limit, we can take advantage of the fact that
the well-measured cosmic star formation rate
is $\csfr = \avg{\psi n_{\rm gal}} \sim \avg{L_{\rm H \alpha} n_{\rm gal}} 
 \sim L_{\star,z} n_{\star,z}$.
This fixes the redshift behavior of the
product $L_{\star,z} n_{\star,z}$, so
that in the limit where one factor is constant,
the other must have the redshift dependence of the cosmic star formation rate.

\subsection{Pure Luminosity Evolution}

In the case of pure luminosity evolution, there is no evolution of the galaxy density. Thus in the Schechter function of eq.~(\ref{eq:lumfunc_S}),
$n_{\star,z} = n_{\star,0}$, and thus all redshift dependence
lies in $\Laz$.  
Therefore, evolution of the star-formation rate in each galaxy, and hence 
the evolution of the galaxy $\rm H{\alpha}$ luminosity $\Laz$, must
trace the general evolution of the cosmic star-formation rate $\csfr$.
Under this assumption, we can show that 
$\avg{M_{\rm gas,S}}$
is independent of the choice of the delay-time function (see derivation in 
Appendix \ref{app:delay}).
When adopting the Schechter function for $\frac{dn_{\rm galaxy, S}}{d\Laz}$,
one will find that $\avg{M_{\rm gas,S}} \propto (1+z)^{-\beta} \ (L_{\star,z})^{\omega} \propto (1+z)^{-\beta} \ (\frac{\csfr(z)}{\csfr(z=0)})^{\omega}$,
with a local value of $\avg{M_{\rm gas,S}}_{z=0} = 6.8 \times 10^9 \ M_{\odot}$ (see Appendix \ref{app:delay}).

\begin{figure}[!h]
\begin{center}
\includegraphics[width=1.0\textwidth]{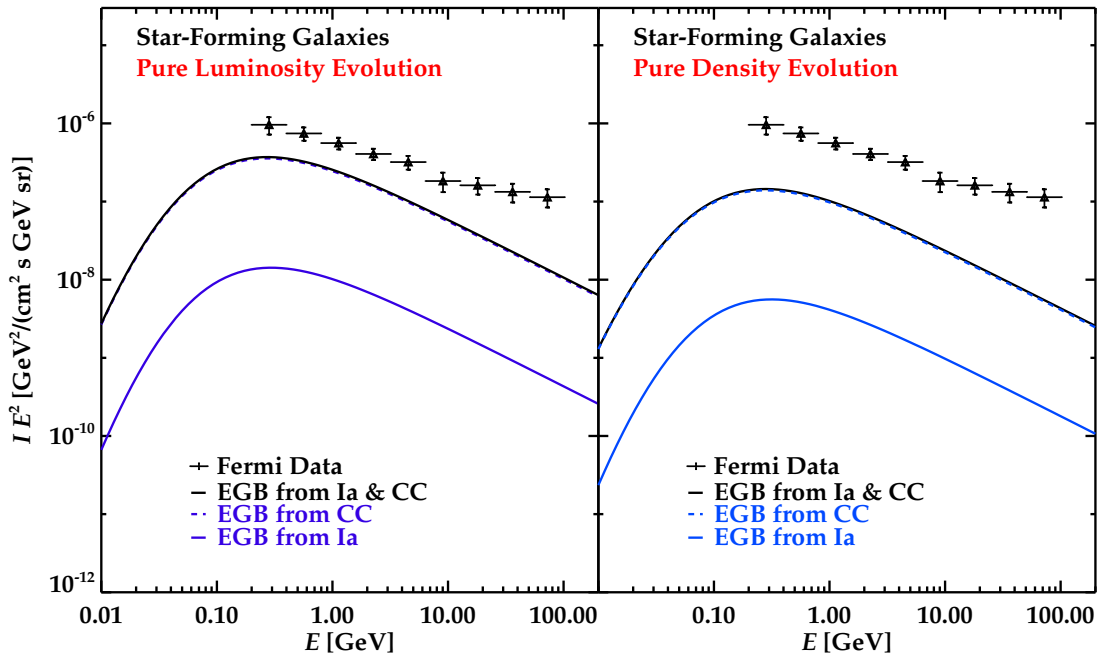}
\end{center}
\caption{
EGB from SNe in star-forming galaxies. Results in the left
panel assume pure luminosity evolution. Results in the 
right panel assume pure density evolution.
The dashed blue line shows the contribution from CC SNe;
the solid blue line shows the contribution from Ia SNe;
and the black line plots the total contribution from both CC
and Ia SNe.
The {\it Fermi} data are obtained from \citet{Abdo09_fermiEGB}.
}
\label{fig:DGRB_S}
\end{figure}

The predicted EGB from Type Ia SNe in star-forming galaxies 
is plotted as the solid blue line in the left panel of Fig.~\ref{fig:DGRB_S}.
For comparison, the dashed blue line shows the EGB contribution
from CC SNe in star-forming galaxies.
The shapes of the dashed blue lines trace the results in \citet{Fields10}.
However, the normalization of the CC SN curves is lower by the fraction of 
the CC SN rate over the total SN rate ($\sim 0.8$ from Bazin et al., 2009),
which is due to the fact that \citet{Fields10} have implicitly
assumed that CC SNe produce all of the gamma-ray emission in galaxies.

Figure~\ref{fig:DGRB_S} shows that the EGB from Type Ia SNe
is around an order of magnitude lower than those
from CC SNe, which is due to the lower Type Ia rate
in star-forming galaxies.
As noted above, this is easily understood as a reflection of
the small Ia/CC ratio (Fig.~\ref{fig:gamma_Ia}).
The black curve in Fig.~\ref{fig:DGRB_S} presents the total EGB emission
from both Type Ia and CC SNe in star-forming galaxies.
Note that  the total EGB emission 
from star-forming galaxies turns out to be very similar to the
prediction in \citet{Fields10}, in which the authors assumed that 
all of the EGB contribution 
comes from the CC events. 
The reason is that even though we added the EGB contribution from Type Ia SNe,
we also lower the EGB emission from CC events estimated 
in \citet{Fields10} by the corresponding 
CC SN fraction. 
Also, the Ia to CC fraction does not change much 
within $z \sim 1$, which is the redshift range where 
most of the EGB signals originate. 
Hence the EGB contribution from CC SNe is always higher than
those from Ia SNe by a similar factor. 

The shape of the EGB curves in Fig.~\ref{fig:DGRB_S} 
traces the general features of the  
pionic gamma-ray energy spectrum.
This is because the observed EGB intensity 
at a specific energy
originated from a combination of sources at different redshifts,
as described in Eq.~(\ref{eq:gamma_general}).
Therefore, the redshift evolution of the unresolved 
sources is smeared out in the energy plot and 
mostly affects the normalization of the EGB intensity 
but not the spectral shape.

\subsection{Pure Density Evolution}
\label{sect:gamma_puredens}

For pure density evolution,
only the galaxy density evolves with redshift
while the galaxy luminosity does not.
Therefore, the star-formation rate $\psi$ in a galaxy
also remains constant,
and the evolution in the cosmic star-formation rate 
will purely depend on the growth 
of the galaxy density.
Hence, in the case of pure density evolution,
$\Lazold = \Laz$.
With similar calculations as those in the case of pure luminosity evolution 
(Appendix \ref{app:delay}),
one can find that $\avg{M_{\rm gas,S}}$ is also independent of 
the choice of the delay-time function.
Additionally, $\avg{M_{\rm gas,S}} \propto (1+z)^{\beta}$
in the case of pure density evolution.

Results for the case of pure density evolution
are shown in the right panel of Fig.~\ref{fig:DGRB_S}.
Again, the solid blue line and the dashed blue line
represent the EGB from Type Ia and CC, respectively.
The black line shows the combined gamma-ray contribution
from both Type Ia and CC events. 
Similar to the results of pure luminosity evolution,
the EGB from Type Ia SNe is lower than that from CC events
because of the lower Type Ia rate.
Moreover, the predicted EGB emission 
is lower if we assume pure density evolution 
instead of pure luminosity evolution.
As discussed in \citep{Fields10}, this is because
a typical galaxy's
gamma-ray luminosity $L_\gamma \propto \psi M_{\rm gas}$;
in the pure luminosity evolution case, both factors are
enhanced at early times, while in the pure density case
this nonlinear boost is not present.

\section{The Extragalactic Gamma-ray Background from Type Ia Supernovae in Quiescent Galaxies}
\label{sect:gamma_Ia_E}

Following a similar procedure to 
\S~\ref{sect:gamma_Ia_S}, we will now discuss the EGB 
from cosmic rays accelerated by Type Ia SNe in quiescent galaxies.
We again express the EGB luminosity density ${\cal L}^{\rm Ia}_{\gamma, \rm Q}$ 
(Eq.~\ref{eq:gamma_L_tot_P_MW}) in the following form to describe the physics inputs from
the average gas mass $\avg{M_{\rm gas, Q}}$ and the cosmic Ia rate in quiescent galaxies $\criap$, 
\beq
\label{eq:gamma_E}
{\cal L}^{\rm Ia}_{\gamma, \rm Q} =  \frac{\epsilon_{\rm Q} \ \criap}
						{\epsilon_{\rm MW} \ R^{\rm MW}_{\rm Ia} + \epsilon_{\rm MW} \ R^{\rm MW}_{\rm CC}} 
 \ \Gamma^{\rm MW}_{\pi^0 \rightarrow \gamma \gamma}(E_{\rm em})
					\ \frac{X_H}{m_p} \
                                \avg{M_{\rm gas,Q}} ,
\eeq
where 
\begin{align}
\avg{M_{\rm gas,Q}} &\equiv \frac{\avg{M_{\rm gas,Q} \ \riap \ n_{\rm galaxy,Q}}}{\avg{\riap \ n_{\rm galaxy,Q}}} \\
                &= \frac{\int dM_{\star, \rm Q} \ M_{\rm gas,Q}(M_{\star, \rm Q}, z) \ \riap(M_{\star, \rm Q},z) \ \frac{dn}{dM_{\star, \rm Q}}}
			{\int dM_{\star, \rm Q} \ \riap(M_{\star, \rm Q},z) \ \frac{dn}{dM_{\star},Q}},
\end{align}
and $\criap \equiv \avg{\riap \ n_{\rm galaxy,Q}}$.
Unlike the star-forming galaxies, where both $\avg{M_{\rm gas,S}}$ and $\crias$
can be related to the observable $\rm H{\alpha}$ luminosity,
it is easier to connect both $\avg{M_{\rm gas,Q}}$ and $\criap$
to the total stellar mass $M_{\star, \rm Q}$ in a quiescent galaxy.

For the cosmic Type Ia rate in quiescent galaxies,
we adopt the results of \citet{Sullivan06},
which link Type Ia rates to $M_{\star, \rm Q}$ directly.
These authors assume a bimodal delay-time distribution and
decompose the Ia rate into two groups: the long-delay time and short-delay time.
In their model, the short-delay time group simply traces the star-formation rate,
while the long-delay time group has a constant probability for all delay times,
i.e., $\rm \Delta(\tau) = \rm constant$. 
Therefore, the Type Ia rate in a galaxy can be written as
\beq
\label{eq:gamma_Iarate_2}
\ria = A \ M_{\star, \rm Q} + B \ \psi.
\eeq
The star formation rate
$\psi \sim 0$ in a quiescent galaxy, thus $\riap = A \ M_{\star, \rm Q}$,
where $M_{\star, \rm Q}$ is the total stellar mass created throughout 
the star formation history in
the quiescent galaxy.
\citet{Sullivan06} estimated $A = 5.1 \times 10^{-14} \ {\rm yr^{-1}} \ M^{-1}_{\odot}$ in
quiescent galaxies based on measurements of the Type Ia rate in the Supernova Legacy Survey (SNLS).

According to the observational results in \citet{Pannella09},
the stellar-mass function $\frac{dn}{dM_{\star},Q}$ of early-type galaxies evolves only slightly with redshift.
Therefore we simply assume the same stellar-mass function throughout all redshift.
Also, we find that the $\frac{dn}{dM_{\star},Q}$ shown in \citet{Pannella09}
can be roughly fitted by the following function
\footnote{
This fitting function is based on observations given in Fig. 7 in \citet{Pannella09}, which contains measurements of 
stellar-mass functions at different redshift bins and environments. We choose the data set 
in the lowest redshift bin of $0.25 <  z < 0.55$
in medium-dense environment 
($\rm log_{10} \rho \sim -2.75$; see \citet{Pannella09} for the definition of environmental density $\rho$ and more details) to
perform a $\chi^2$ fitting.
The reduced-$\chi^2$ of our fitting function is $\sim 0.22$.
},
\beq
\label{eq:gamma_MF}
\frac{dn}{d (\log_{10} M_{\star, \rm Q})} = C_m \ 
  \exp\left( -\frac{\log_{10}^2 (M_{\star, \rm Q}/\mu)}{\sigma^2_m} \right)
\eeq 
where $C_m = 2.05 \times 10^{-3} \ {\rm Mpc^{-3} \ log^{-1}} \ M_{\odot}$, 
$\mu = 10^{10.7} M_\odot$, and $\sigma_m = 0.77$.
Equation~(\ref{eq:gamma_MF}) (the stellar mass function for quiescent galaxies) 
and Eq.~(\ref{eq:gamma_Iarate_2}) (Type Ia rate in a quiescent galaxy)
give the cosmic Ia rate in quiescent galaxies
\begin{align*}
\criap &= \int dM_{\star, \rm Q} \ \riap \ \frac{dn}{dM_{\star, \rm Q}} \\
                	       &= 1.53 \times 10^{-5} \ \rm yr^{-1} \ \rm Mpc^{-3}.
\end{align*}
Here the constant rate follows from the nearly redshift-independent
quiescent stellar mass function.

Most of the gas content in quiescent galaxies appears to be in the form
of diffuse hot gas and can be observed in the X-ray \citep[e.g.,][]{Forman85,Canizares87,Bregman92}.
However, large uncertainties exist in estimations of the mass of hot
gas.
Some studies suggest that most of the quiescent galaxies are gas-poor \citep[e.g.,][]{David06,Fukazawa06},
while other studies imply that there can be significant amount of gas 
in these galaxies \citep[e.g,][]{Jiang07,Humphrey11}.
Current theoretical models suggest that the gas 
could extend out to a few hundred kpc from the center.
The total gas mass of a single galaxy can increase by
orders of magnitude if one includes the gas at large radii \citep{Humphrey10,Humphrey11} 
Therefore observational measures of gas mass can vary 
even for a single galaxy, depending on whether or not 
the observations enclose a large enough radius to include 
gas at large distances.
If hot gas extends to large
radii, this could serve as a reservoir of targets
for cosmic rays accelerated by Type Ia SNe,
provided the cosmic rays also reach large radii.
This too is uncertain, and will depend on the 
cosmic-ray confinement versus escape properties of these
galaxies.

Because of the large uncertainties in the gas content in quiescent galaxies,
we illustrate a range of EGB predictions for these objects, based 
on three different gas amounts.
Since 
quiescent galaxies are dominantly early-type,
we will use estimates of the gas mass in early-type galaxies 
for the amount of gas in quiescent galaxies.
The stellar-mass function in \citet{Pannella09} (Eq.~\ref{eq:gamma_MF})
gives the average stellar mass in early-type 
galaxies to be $\avg{M_{\star, \rm Q}} = 5.17 \times 10^{11} \ M_{\odot}$,
which can be converted to the total gas mass 
by multiplying a gas-to-stellar mass ratio $M_{\rm gas}/M_{\star}$, i.e.,
$\avg{M_{\rm gas, Q}} = (M_{\rm gas, Q}/M_{\star, \rm Q}) \ \avg{M_{\star, \rm Q}}$.
The three models we adopted for different gas amounts correspond to 
different gas-to-stellar mass ratios $M_{\rm gas, Q}/M_{\star, \rm Q}$ for early-type galaxies.
Table~\ref{tab:gamma_gas} summarizes the gas-to-stellar mass ratios $M_{\rm gas, Q}/M_{\star, \rm Q}$
for the three models we adopted, as well as the corresponding $\avg{M_{\rm gas, Q}}$.
Gas Model 1 estimates the gas-to-stellar mass ratio based on the stellar-mass fraction of the total halo mass from \citet{Jiang07} 
\footnote{
\citet{Jiang07} found that the average stellar-mass fraction of the total halo mass
in early-type galaxies is $M_{\star}/M_{\rm tot} \sim 0.026$ or $0.056$ based on different
assumptions of the halo mass dynamics. Both of these numbers are significantly lower
than the cosmological baryon-to-mass ratio $\Omega_b / \Omega_m \sim 0.176$
measured by {\it WMAP} \citep{wmap07}. If we assume that the baryon-to-mass ratio in a galaxy
can be well represented by the cosmological ratio, i.e., $M_b/M_{\rm tot} \sim \Omega_b / \Omega_m$,
the result from \citet{Jiang07} implies a large amount of gas mass in
early-type galaxies, which can be estimated by
$M_{\rm gas, Q} = (M_{\rm baryon, Q} - M_{\star, \rm Q}) \sim M_{\star, \rm Q} \ (\frac{\Omega_b / \Omega_m}{M_{\star} / M_{\rm tot}} - 1)$.
The values of $M_{\star}/M_{\rm tot} \sim 0.026$ and $0.056$ correspond to
$M_{\rm gas, Q} = 5.77 \ M_{\star, \rm Q}$ and $M_{\rm gas, Q} = 2.14 \ M_{\star, \rm Q}$, respectively.
Here we adopt the latter number to be more conservative in our estimation.
}.
This model gives the highest gas-to-stellar mass ratio of all three models. 
Gas Model 2 and 3 are both adopted from \citet{David06}, in which the authors reported
the gas-to-stellar mass ratio for more luminous (Gas Model 2) and less luminous (Gas Model 3)
early-type galaxies. 
Note that both of the gas mass and the Ia rate in a quiescent galaxy are constant with redshift, as a result
of assuming a non-evolving stellar-mass function, consistent with the early-type galaxy observations in \citet{Pannella09}.

\begin{table}[ht]
\caption{\label{tab:gamma_gas}
Summary of different gas models adopted in the EGB calculations.}
\begin{center}
\begin{tabular}{|c||c|c|c|c|}
\hline\hline
Gas Model  & $M_{\rm gas, Q}/M_{\star, \rm Q}$ & $\avg{M_{\rm gas,Q}}$ & Reference \\
\hline
1 & 2.14 & $1.11 \times 10^{12} \ M_{\odot}$ & \citet{Jiang07} \\
2 & 0.01& $5.17 \times 10^{9} \ M_{\odot}$ & \citet{David06} \\
3 & 0.001 & $5.17 \times 10^{8} \ M_{\odot}$ & \citet{David06} \\
\hline\hline
\end{tabular}
\end{center}
\end{table}

\begin{figure}[!h]
\begin{center}
\includegraphics[width=1.0\textwidth]{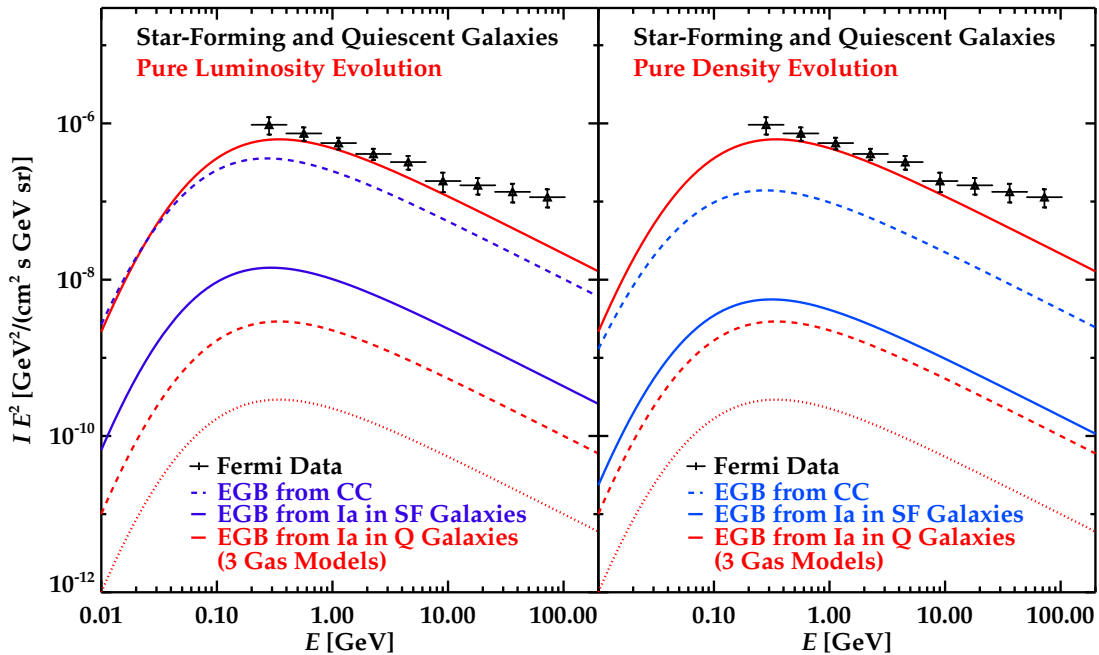}
\end{center}
\caption{
EGB from SNe in both star-forming and quiescent galaxies.
Results in the left panel assume pure luminosity evolution for star-forming galaxies.
Results in the right panel assume pure density evolution for star-forming galaxies.
The dashed blue line shows the contribution from CC SNe;
the solid blue line shows the contribution from Ia SNe
in star-forming galaxies;
the red lines show the contribution from Ia SNe
in quiescent galaxies based on different gas models
(solid red: Gas Model 1, dashed red: Gas Model 2, dotted red: Gas Model 3).
The {\it Fermi} data are obtained from \citet{Abdo09_fermiEGB}.
}
\label{fig:DGRB_E}
\end{figure}

\begin{figure}[!h]
\begin{center}
\includegraphics[width=1.0\textwidth]{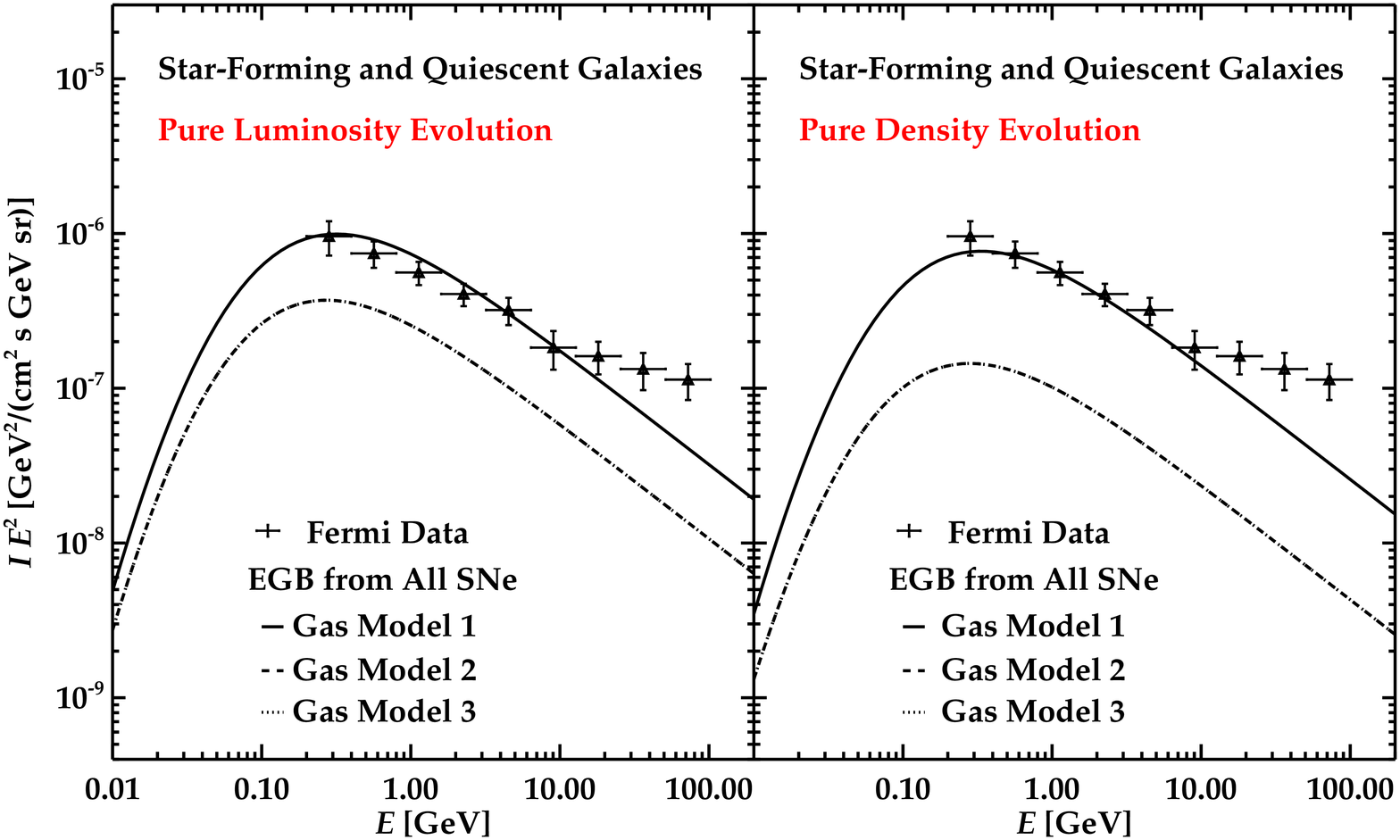}
\end{center}
\caption{
Total EGB from Ia and CC SNe 
in both star-forming and quiescent galaxies.
Results in the left panel assume pure luminosity evolution.
Results in the right panel assume pure density evolution.
The {\it Fermi} data are obtained from \citet{Abdo09_fermiEGB}.
}
\label{fig:DGRB_total}
\end{figure}

The red curves in Fig.~\ref{fig:DGRB_E} 
plot the EGB estimation from Type Ia SNe in quiescent galaxies,
with different line styles correspond to estimations from different gas amounts
(solid line: Gas Model 1, dashed line: Gas Model 2, dotted line: Gas Model 3).
The EGB emissions from SNe in star-forming galaxies are plotted as blue curves
for comparison.
The black curves in Fig.~\ref{fig:DGRB_total} plot the total EGB emissions from SNe in both star-forming and quiescent galaxies
for different gas models.
For both Fig.~\ref{fig:DGRB_E}  and ~\ref{fig:DGRB_total},
the left panel 
plots results under the assumption of pure luminosity evolution
for the star-forming galaxies.
The right panel shows the EGB predictions assuming pure density evolution
for the quiescent galaxies.

The estimated EGBs shown in Fig.~\ref{fig:DGRB_E} are linearly proportional to the average galaxy gas mass $\avg{M_{\rm gas,Q}}$ and thus
to the adopted gas-to-stellar mass ratio, 
as required by Eq.~(\ref{eq:gamma_E}).
Gas Model 1 predicts the highest contribution in EGB from quiescent galaxies,
where the result
is significantly higher than the EGB emission from Type Ia events in star-forming galaxies,
while Gas Model 2 and Gas Model 3 suggest much lower EGB emission from quiescent galaxies.
In our estimation, the two important factors that affect the overall EGB are the average gas mass and the cosmic SN rate.
In general, the cosmic Type Ia rate is about a factor of $5 -10$ smaller than the cosmic CC SN rate. 
Additionally, the cosmic Type Ia rate in all quiescent galaxies 
averaging over the entire redshift range is lower than that in 
star-forming galaxies by around a factor of three.

As for the gas mass in quiescent galaxies, Gas Model 1 assumes 
the gas amount to be about two orders of magnitude higher than that in star-forming galaxies,
while Gas Model 2 and 3 assume gas amount to be similar or one order of magnitude lower than 
that in star-forming galaxies, respectively.
Combining these two factors (gas mass and Type Ia rate),
we would expect the EGB from quiescent galaxies to be 
about 30 times larger (for Gas Model 1), 3 times smaller (for Gas Model 2), or 30 times smaller (for Gas Model 3) 
than that from Type Ia in star-forming galaxies.
Because of the EGB estimation is very sensitive to the gas amount,
measurement of the 
EGB could put constraints on the gas mass in quiescent galaxies.

\section{The Uncertainties in the Extragalactic Gamma-ray Background Analysis}

\label{sect:gamma_uncertainty}

\subsection{Star-forming Galaxies}

The gamma-ray emission from normal star-forming galaxies
is the most constrained among the three galaxy classes that
are considered possible SN-induced EGB sources (i.e., starburst galaxies, 
normal star-forming galaxies, and quiescent galaxies).
The main uncertainties in the EGB prediction of star-forming galaxies come
from four factors, as described in \citet{Fields10}:
(1) uncertainty in the pionic gamma-ray production rate
$\Gamma^{\rm MW}_{\pi^0 \rightarrow \gamma \gamma}(E_{\rm em})$,
which is $\sim 30\%$ \citep{Abdo09_CRint},
(2) uncertainty in the normalization of the Galactic SN rate $R^{\rm MW}
_{\rm SN}$,
which is $\sim 40\%$ \citep{Robitaille10},
(3) uncertainty in the luminosity scaling in $\avg{M_{\rm gas,S}}$,
which is $\sim 25\%$ \citep{Fields10}, and
(4) uncertainty in the normalization of the cosmic SN rate ${\cal R}_{\rm SN,S}$,
which is $\sim 16\%$ resulting from the uncertainties 
in the cosmic CC SN rate ${\cal R}_{\rm CC} \sim (1.0 \pm 0.2) \times 10^{-4} \ \rm yr^{-1} \ Mpc^{-3}$ \citep{Horiuchi09}
and the cosmic Ia rate ${\cal R}_{\rm Ia} \sim (0.25 \pm 0.05) \times 10^{-4} \ \rm yr^{-1} \ Mpc^{-3}$ \citep{Horiuchi10}.
The total uncertainty in the EGB prediction will then be $\sim 10^{\pm 0.25}$.

The upcoming large synoptic surveys,
such as the LSST, will provide novel information in both the cosmic SN rate
and how they depend on the galaxy types out to high redshift.
Within one year of observation, LSST is expected to detect $\sim 10^5$ SNe out to $z \sim 1$
and thus achieve
a statistical precision of less than a few percent in the cosmic SN rate \citep{lf, LSSTsb_Ia}.
Hence, LSST will essentially remove the uncertainty from the cosmic SN rate
in the EGB analysis 
and make the EGB a better tool for studying cosmic rays and gamma-ray physics. 
Moreover, such a large SN population will provide excellent statistics
for study of how SN rates evolve as a function of 
SN type (and sub-type), host galaxy type and star-formation rate,
and cosmic environment.  Thus it will be possible to directly measure the
Type Ia rate in quiescent galaxies and its evolution with redshift.

\subsection{Quiescent Galaxies: Hot Gas and Cosmic-Ray Propagation}

\label{sect:gamma_uncertainty_Q}

For quiescent galaxies, many characteristics related to their gamma-ray emissions
are poorly understood.
The Type Ia SN rate in these systems could be uncertain
up to a factor $\sim 2$, particularly at $z > 1$.
We have also seen that the gas mass in these systems
is even more poorly known.
Published estimates of the gas content in
quiescent galaxies vary by orders of magnitude.
Observations that use X-rays as tracers of hot gas
found significantly less gas in quiescent galaxies \citep[e.g.,][]{David06,Fukazawa06}.
However,
several studies imply a much higher gas content in quiescent galaxies 
than previously thought, when using gravitational lensing and
modeling that involved dark matter \citep[e.g.,][]{Jiang07,Capelo10,Humphrey11}.
It is difficult to determine what are the causes of the 
large discrepancy of gas quantity from different analyses,
because these studies adopted different observational and modeling techniques
with different galaxy sample. 
One possible reason for the large variation in gas content measurements  
might 
come from whether or not one includes gas at larger radii, which are 
predicted by some theoretical models \citep{Capelo10,Humphrey10,Humphrey11}.

If gas content does extend to larger radii in quiescent galaxies,
it is crucial to understand the cosmic-ray propagation in quiescent galaxies 
to determine the confinement volume of cosmic rays and how likely they 
can interact with gas at larger radii.
Unfortunately, cosmic-ray propagation remains poorly understood.
Until now, most of the studies have been focused on the Milky Way or spiral galaxies and 
not so much on quiescent galaxies.
For spiral galaxies, both observations and theoretical modeling suggest that 
most of the cosmic rays are confined within $\sim$ kpc of the disk \citep{Stecker77, Strong00,Strong04}.
However, this might not be the case for quiescent galaxies.
\citet{Hein08} simulate cosmic-ray propagation in elliptical galaxies with the presence of diffusion, as well as adiabatic losses.
They apply their formalism not to SN sources but rather to acceleration due to an M87-like relativistic jet; 
thus their detailed calculations are not applicable to our case.   Nevertheless, their general finding is that cosmic rays expand
into a larger volume in an elliptical galaxy than in a spiral galaxy, due to the larger minor axis in an elliptical galaxy.
Additionally, they argue that adiabatic losses are much more important in elliptical galaxies and indeed dominate over escape losses.
\citet{Hein08} thus conclude that it is likely for elliptical galaxies 
to be extended gamma-ray sources. 

There are two factors that are important for determining the probability of interactions between cosmic rays 
and gas, especially at large radii where the density of gas and cosmic rays are likely to be smaller: 
(1) the mean free time for pion production for each cosmic-ray particle 
$\tau_{\rm pp \rightarrow \pi^0} = (n_{\rm gas} \sigma_{\rm pp \rightarrow \pi^0} c)^{-1} \sim 1 \ {\rm Gyr} \ (0.1 \ {\rm cm^2}/n_{\rm gas})$, 
where $\sigma_{\rm pp \rightarrow \pi^0}$ is the cross section of pion production and $n_{\rm gas}$ is the number density of gas particles,
and (2) time $\tau_{\rm esc}$ that a cosmic ray particle takes
to propagate through the galaxy before escape. 
In the limit of cosmic rays travel through a (non-magnetized) galaxy radially, that is,
without any diffusion, the escape time is very short and
probability of interactions between cosmic rays and gas particles is small.
On the other hand, in the presence of disordered magnetic fields
extending to large radii,
cosmic ray propagation 
will not be radial but diffusive instead, which will make the escape time much larger
and increase the chance of cosmic-ray interactions with ISM.

Clearly there are fundamental uncertainties (and opportunities) in a realistic treatment of elliptical galaxy cosmic rays, but
even in our simplistic picture the parameters $\Lambda_{\rm esc}$ and $\epsilon_{\rm Q}$ are not well-constrained.
Although we treated these two quantities to be the same for both Type Ia 
and CC SNe in all environments due to limited knowledge, it is possible, and even likely, 
that these numbers
are different in quiescent galaxies.
In fact,
\citet{Dorfi96} have suggested
that the efficiency in quiescent galaxies is at least 10 times lower than
that in star-forming galaxies \citep{Dorfi96},
which could lower our prediction of the EGB from quiescent galaxies
by a factor of 10 or even larger.
\citet{Tang05} model the evolution of SN remnants in low-density hot media
and reach similar conclusion that the SN heating in such an environment is subtle 
and cover a large region because of small Mach numbers. 
Likewise, a smaller escape path length, i.e., a weaker cosmic-ray confinement, 
can also decrease our EGB estimation in quiescent galaxies.
Note that adopting different values of $\epsilon$ and $\Lambda_{\rm esc}$ would change the cosmic-ray spectrum,
and hence also change the corresponding gamma-ray spectrum.

To summarize, current studies imply that cosmic-ray acceleration might be more difficult in quiescent galaxies 
than in star-forming galaxies due to weaker shocks in low-density and high-temperature environments \citep{Dorfi96, Tang05}. 
However, quiescent galaxies might have larger confinement volume 
than spiral galaxies because they have larger semi-minor radii \citep{Hein08}. 
Further observational and theoretical study of the gas and cosmic-ray content of quiescent galaxies
clearly are needed in order to pin down the gamma-ray production of
these objects even to within an order of magnitude.

\section{Conclusions}

\label{sect:gamma_conclusion}

We have calculated the EGB contribution from Type Ia SNe 
in both star-forming and quiescent galaxies,
extending the work of \citet{Fields10}.
For star-forming galaxies, most of the 
gamma-ray emission comes from cosmic rays 
accelerated by CC SNe. This is mainly 
because there are about five times more 
CC SNe than Type Ia events 
in star-forming galaxies.
We find that the net EGB contribution from both SN
types is almost the same as the 
\citet{Fields10} predictions that 
only included CC events.
Our model allows for addition of cosmic Type Ia explosions, but
also includes these in the
cosmic-ray/star-formation ratio, which we normalize to the Milky
Way values.  Both factors change by nearly the same amount,
so that the addition of Type Ia events causes almost no net change
in the EGB prediction.

We also point out that Type Ia events
in quiescent galaxies
make a unique contribution to the EGB,
because these systems lack CC events.
We show that
the EGB from Type Ia events in quiescent galaxies
is highly sensitive to the gas amount in quiescent galaxies,
which is still poorly known.
Based on different gas models adopted,
the EGB from Type Ia SNe in quiescent galaxies
can vary from two orders of magnitudes higher to an order of magnitude lower than
those produced by Type Ia SNe in star-forming galaxies.
Therefore quiescent galaxies can be dominant source of the EGB 
if there exist a large amount of gas, as suggested by \citet{Jiang07} and \citet{Humphrey11}.
The measurement of the EGB will provide useful constraint on the gas amount in quiescent galaxies.
Additional uncertainties 
in the cosmic-ray acceleration efficiency and confinement
could also change the 
EGB emission in these systems by a few orders of magnitude.
Hence the EGB can also provide limits on the 
these two quantities in 
quiescent galaxies.

It is thus important to understand the 
characteristics of cosmic SNe of all types,
in order to 
correctly predict their contribution to 
the EGB.
We conclude that the large 
SN sample provided by LSST will
offer critical information about the cosmic SN 
rate for both CC and Ia events, 
and their dependence on galaxy types out to high redshift.

The {\it Fermi} detection of the EGB contains
crucial information about the extragalactic 
gamma-ray source spectrum. Particularly,
it can provide an important probe 
to the gas amount, the cosmic-ray acceleration efficiency
and the cosmic-ray confinement 
in quiescent galaxies. With our knowledge 
about SNe increasing rapidly 
as future synoptic surveys come online,
the EGB contribution from SNe in galaxies
can possibly be disentangled from other source candidates.

{\bf Acknowledgments}
We are thankful for the helpful comments and suggestions from Floyd Stecker that
greatly improved this paper.
We also appreciate the helpful discussions with Tonia Venters, Don Ellison, Robert Brunner, 
Tijana Prodanovic, Vasiliki Pavlidou,
Ann Hornschemeier, Bret Lehmer, and Theresa Brandt.
Additionally, we are grateful for the thoughtful comments from the anonymous referee.
This work was partially supported by NASA via the Astrophysics
Theory Program through award NNX10AC86G.

\appendix

\section{The Delay-Time Distribution of Type Ia Supernovae and Detailed Calculation of $\avg{M_{\rm gas, S}}$ in Equation~\ref{eq:gamma_Ia}}

\label{app:delay}

The delay time of each Type Ia SN can differ from $\sim 0.1$ Gyr to $\sim 10$ Gyr 
\citep{Mannucci05, Scannapieco05, gal-yam,Mannucci06, Sullivan06, Maoz11}. 
Observationally, the delay times of Type Ia SNe are usually studied via comparison between 
measurements of the cosmic Type Ia SN rate and the cosmic star-formation rate,
and are usually described by some functions of delay-time distribution,
which describes the probability of a Type Ia event with a specific
delay time.
Current proposed delay-time distributions have included
a single power law \citep[e.g.,][]{Horiuchi10,Graur11},
a Gaussian \citep[e.g.,][]{Strolger04, Dahlen08_Ia},
and a bimodal distribution \citep[e.g.,][]{Sullivan06}.
The difficulty in determining the delay-time distribution 
mainly comes from the uncertainty in the cosmic Ia SN rate measurements.
Fortunately, the value $\avg{M_{\rm gas, S}}$ is independent of the 
choice of delay-time distribution, as we show in the following derivation.

Based on the relation between the gas mass $M_{\rm gas, S}$ and the star-formation rate $\psi$ (Eq.~\ref{eq:M_gas_S}),
and also the connection between $\psi$ and $\Laz$, i.e., $\psi(\gammaLa, z)/(1 \ M_{\odot} \ \rm yr^{-1}) = \Laz / (1.26 \times 10^{34} \ \rm W)$,
the $M_{\rm gas,S}$ for a galaxy at redshift $z$ can be directly linked to the observable $H_{\alpha}$ luminosity $\Laz$
by $M_{\rm gas,S} = 2.8 \times 10^9 \ M_{\odot} \ (1+z)^{-\beta} \ (\frac{\Laz}{1.26 \times 10^{34} \ \rm W})^{\omega}$.
Additionally,
the Type Ia rate can also be related to $\Laz$ by
\beq
\rias \propto \int^t_0 L_{H_{\alpha}}(t-\tau) \ \rm \Delta(\tau) \ d\tau,
\eeq
where $L_{H_{\alpha}}(t-\tau) \equiv \Lazold(t,\tau)$, which is the $H_{\alpha}$ luminosity
measured at some earlier time $t-\tau$ or larger redshift $z+\Delta z$.
Therefore $\avg{M_{\rm gas,S}}$ can be expressed in terms of $\Laz$,
\beq
\label{eq:gamma_M_Ias}
\avg{M_{\rm gas,S}} \propto \frac{\int d\Laz \ (1+z)^{-\beta} \ (\Laz)^{\omega} \
                        (\int^t_0 \ \Lazold(t,\tau) \ \rm \Delta(\tau) \ d\tau)
                        \frac{dn_{\rm galaxy, S}}{d\Laz}}
                {\int d\Laz \ (\int^t_0 \ \Lazold(t,\tau) \ \rm \Delta(\tau) \ d\tau) \ \frac{dn_{\rm galaxy, S}}{d\Laz}}
\eeq
This equation expresses only the redshift-dependent terms 
and we will further calculate how $\avg{M_{\rm gas,S}}$ evolves with redshift
in the case of pure luminosity evolution and pure density evolution, respectively, in 
\ref{sect:gamma_purelume_A} and \ref{sect:gamma_puredens_A}.
\citet{Fields10} shows that the local value of $\avg{M_{\rm gas,S}}_{z=0} = 6.8 \times 10^9 \ M_{\odot}$.

\subsection{Pure Luminosity Evolution}
\label{sect:gamma_purelume_A}

In the case of pure luminosity evolution,
the star-formation rate in each galaxy traces the general evolution 
of the cosmic star-formation rate,
i.e., $\frac{\psi(z+\Delta z)}{\psi(z)} = \frac{\csfr(z+\Delta z)}{\csfr(z)}$.
Therefore from the directly proportional relation between the 
star-formation rate $\psi$ in a galaxy and the galaxy $H_{\alpha}$ luminosity $\Laz$,
one can trace the evolution of the $H_{\alpha}$ luminosity via 
the history of cosmic star-formation rate,
which is well-known out to redshift $z \sim 1$
\citep[e.g.,][and references therein]{Hopkins04, Hopkins06}.
That is,
$\frac{\Lazold}{\Laz} = \frac{\psi(z+\Delta z)}{\psi(z)} = \frac{\csfr(z+\Delta z)}{\csfr(z)}$.
Therefore the galaxy luminosity at different redshifts can be found by
\beq
\label{eq:gamma_lumtocsfr}
\Lazold = \Laz \ \frac{\csfr(z+\Delta z)}{\csfr(z)} \equiv \Laz \ \frac{\csfr(t-\tau)}{\csfr(t)} .
\eeq
With this relation,
$\avg{M_{\rm gas,S}}$ in Eq.~(\ref{eq:gamma_M_Ias}) can be simplified to
\begin{align}
\label{eq:gamma_M_gas_S_L}
\avg{M_{\rm gas,S}}      
                &\propto \frac{\int d\Laz (1+z)^{-\beta} \ (\Laz)^{\omega} \ 
                                (\int^t_0 \ \Laz \ \frac{\csfr(t-\tau)}{\csfr(t)} \ \Delta(\tau) \ d\tau) 
                                \frac{dn_{\rm galaxy, S}}{d\Laz}}
                        {\int d\Laz \ (\int^t_0 \ \Laz \ \frac{\csfr(t-\tau)}{\csfr(t)} \ \Delta(\tau) \ d\tau) \ \frac{dn_{\rm galaxy, S}}{d\Laz}} \\
                &= \frac{(1+z)^{-\beta} \ (\int^t_0 \ \frac{\csfr(t-\tau)}{\csfr(t)} \ \Delta(\tau) \ d\tau) \                        
                                \int d\Laz \ (\Laz)^{\omega+1} \ \frac{dn_{\rm galaxy, S}}{d\Laz}}                        
                        {(\int^t_0 \ \frac{\csfr(t-\tau)}{\csfr(t)} \ \Delta(\tau) \ d\tau) \                         
                                \int d\Laz \ \Laz \ \frac{dn_{\rm galaxy, S}}{d\Laz}} \\                
                &= \frac{(1+z)^{-\beta} \
                        \int d\Laz \ (\Laz)^{\omega+1} \ \frac{dn_{\rm galaxy, S}}{d\Laz}}
                        {\int d\Laz \ \Laz \ \frac{dn_{\rm galaxy, S}}{d\Laz}}.
\end{align}
Therefore $\avg{M_{\rm gas,S}}$ is independent of the choice of delay-time distribution.

The assumption of pure luminosity evolution implies that $n_{\star,z}$ in the Schechter function remains constant 
and $L_{\star,z}$ in the Schechter function
evolves as $\csfr$ (Eq.~\ref{eq:gamma_lumtocsfr}).
Hence the redshift dependence of $\avg{M_{\rm gas, S}}$
can be further calculated using the Schechter function,
\begin{align}
\avg{M_{\rm gas,S}}  &\propto (1+z)^{-\beta} \ L^{\omega}_{\star,z} \ \frac{
                        \int^{L_{\rm max}} d(\frac{\Laz}{L_{\star,z}}) \ (\frac{\Laz}{L_{\star,z}})^{\omega+1} \ \frac{n_{\star,z}}{L_{\star,z}} \ (\frac{\Laz}{L_{\star,z}})^{-\alpha} \ e^{-\Laz/L_{\star,z}}}
                        {\int^{L_{\rm max}} d(\frac{\Laz}{L_{\star,z}}) \ \frac{\Laz}{L_{\star,z}} \ \frac{n_{\star,z}}{L_{\star,z}} \ (\frac{\Laz}{L_{\star,z}})^{-\alpha} \ e^{-\Laz/L_{\star,z}}} \\
		      &= (1+z)^{-\beta} \ L^{\omega}_{\star,z=0} \ (\frac{\csfr(z)}{\csfr(z=0)})^{\omega} \ \frac{
                        \int^{L_{\rm max}} d(\frac{\Laz}{L_{\star,z}}) \ (\frac{\Laz}{L_{\star,z}})^{\omega+1} \ \frac{n_{\star,z}}{L_{\star,z}} \ (\frac{\Laz}{L_{\star,z}})^{-\alpha} \ e^{-\Laz/L_{\star,z}}}
                        {\int^{L_{\rm max}} d(\frac{\Laz}{L_{\star,z}}) \ \frac{\Laz}{L_{\star,z}} \ \frac{n_{\star,z}}{L_{\star,z}} \ (\frac{\Laz}{L_{\star,z}})^{-\alpha} \ e^{-\Laz/L_{\star,z}}} \\
		      &\propto (1+z)^{-\beta} \ (\frac{\csfr(z)}{\csfr(z=0)})^{\omega} 
\end{align}
where $L_{\rm max}$ is the maximum luminosity for star-forming galaxies, which corresponds to the maximum star formation defined in \citet{Fields10}.
Galaxies with luminosities greater than $L_{\rm max}$ are considered starburst galaxies and are not included in this calculation. 
Additionally, we adopt the cosmic star-formation rate $\csfr$ described in \citet{Horiuchi09} based on current observations.
Note that because the factors related to delay-time distribution canceled out, 
this result turns out to be the same as the one obtained in \citet{Fields10}.

\subsection{Pure Density Evolution}
\label{sect:gamma_puredens_A}

In the case of pure density evolution $\Lazold = \Laz$ as discussed in \S~\ref{sect:gamma_puredens}.
Thus in the Schechter function, $L_{\star,z}$ remains constant while $n_{\star,z}$ evolves as $\csfr$.
With similar calculations shown in the case of pure luminosity evolution (\ref{sect:gamma_purelume_A}) and adopting the Schechter function for 
$\frac{dn_{\rm galaxy, S}}{d\Laz}$, we can derive the redshift evolution of $\avg{M_{\rm gas,S}}$
in the case of pure density evolution:
\begin{align}
\avg{M_{\rm gas,S}}     
                &\propto \frac{\int d\Laz (1+z)^{-\beta} \ (\Laz)^{\omega} \ 
                                (\int^t_0 \ \Laz \ \Delta(\tau) \ d\tau) 
                                \frac{dn_{\rm galaxy, S}}{d\Laz}}
                        {\int d\Laz \ (\int^t_0 \ \Laz \ \ \Delta(\tau) \ d\tau) \ \frac{dn_{\rm galaxy, S}}{d\Laz}} \\
                &= \frac{(1+z)^{-\beta} \ (\int^t_0 \ \Delta(\tau) \ d\tau) \                        
                                \int d\Laz \ (\Laz)^{\omega+1} \ \frac{dn_{\rm galaxy, S}}{d\Laz}}                        
                        {(\int^t_0 \ \Delta(\tau) \ d\tau) \                         
                                \int d\Laz \ \Laz \ \frac{dn_{\rm galaxy, S}}{d\Laz}} \\                
                &= \frac{(1+z)^{-\beta} \
                        \int d\Laz \ (\Laz)^{\omega+1} \ \frac{dn_{\rm galaxy, S}}{d\Laz}}
                        {\int d\Laz \ \Laz \ \frac{dn_{\rm galaxy, S}}{d\Laz}} \\
		&\propto (1+z)^{-\beta}.	       
\end{align}
Again, because $\avg{M_{\rm gas,S}}$ is independent of the choice of delay-time distribution,
the result is identical to the one calculated in \citet{Fields10}.

\bibliographystyle{apj}
\bibliography{ref_thesis}

\begin{thebibliography}{100}
\expandafter\ifx\csname natexlab\endcsname\relax\def\natexlab#1{#1}\fi

\bibitem[{{Abdo} {et~al.}(2009{\natexlab{a}}){Abdo}, {Ackermann}, {Ajello},
  {Anderson}, {Atwood}, {Axelsson}, {Baldini}, {Ballet}, {Barbiellini},
  {Bastieri}, {Baughman}, {Bechtol}, {Bellazzini}, {Berenji}, {Blandford},
  {Bloom}, {Bonamente}, {Borgland}, {Bregeon}, {Brez}, {Brigida}, {Bruel},
  {Burnett}, {Caliandro}, {Cameron}, {Caraveo}, {Casandjian}, {Cecchi},
  {Charles}, {Chekhtman}, {Cheung}, {Chiang}, {Ciprini}, {Claus},
  {Cohen-Tanugi}, {Conrad}, {Dereli}, {Dermer}, {de Angelis}, {de Palma},
  {Digel}, {di Bernardo}, {Dormody}, {Do Couto E Silva}, {Drell}, {Dubois},
  {Dumora}, {Edmonds}, {Farnier}, {Favuzzi}, {Fegan}, {Focke}, {Frailis},
  {Fukazawa}, {Funk}, {Fusco}, {Gaggero}, {Gargano}, {Gehrels}, {Germani},
  {Giebels}, {Giglietto}, {Giordano}, {Glanzman}, {Godfrey}, {Grenier},
  {Grondin}, {Grove}, {Guillemot}, {Guiriec}, {Hanabata}, {Harding},
  {Hayashida}, {Hays}, {Hughes}, {J{\'o}hannesson}, {Johnson}, {Johnson},
  {Johnson}, {Johnson}, {Kamae}, {Katagiri}, {Kataoka}, {Kawai}, {Kerr},
  {Kn{\"o}dlseder}, {Kocian}, {Kuehn}, {Kuss}, {Lande}, {Latronico}, {Longo},
  {Loparco}, {Lott}, {Lovellette}, {Lubrano}, {Madejski}, {Makeev},
  {Mazziotta}, {McConville}, {McEnery}, {Meurer}, {Michelson}, {Mitthumsiri},
  {Mizuno}, {Moiseev}, {Monte}, {Monzani}, {Morselli}, {Moskalenko}, {Murgia},
  {Nolan}, {Nuss}, {Ohsugi}, {Okumura}, {Omodei}, {Orlando}, {Ormes},
  {Paneque}, {Panetta}, {Parent}, {Pelassa}, {Pepe}, {Pesce-Rollins}, {Piron},
  {Porter}, {Rain{\`o}}, {Rando}, {Razzano}, {Reimer}, {Reimer}, {Reposeur},
  {Ritz}, {Rodriguez}, {Roth}, {Ryde}, {Sadrozinski}, {Sanchez}, {Sander}, {Saz
  Parkinson}, {Scargle}, {Sellerholm}, {Sgr{\`o}}, {Smith}, {Smith}, {Spandre},
  {Spinelli}, {Starck}, {Stecker}, {Striani}, {Strickman}, {Strong}, {Suson},
  {Tajima}, {Takahashi}, {Tanaka}, {Thayer}, {Thayer}, {Thompson}, {Tibaldo},
  {Torres}, {Tosti}, {Tramacere}, {Uchiyama}, {Usher}, {Vasileiou}, {Vilchez},
  {Vitale}, {Waite}, {Wang}, {Winer}, {Wood}, {Ylinen}, \&
  {Ziegler}}]{Abdo09_fermiEGB}
{Abdo}, A.~A., {et~al.} 2009{\natexlab{a}}, Physical Review Letters, 103,
  251101

\bibitem[{{Abdo} {et~al.}(2009{\natexlab{b}}){Abdo}, {Ackermann}, {Ajello},
  {Baldini}, {Ballet}, {Barbiellini}, {Baring}, {Bastieri}, {Baughman},
  {Bechtol}, {Bellazzini}, {Berenji}, {Blandford}, {Bloom}, {Bonamente},
  {Borgland}, {Bouvier}, {Bregeon}, {Brez}, {Brigida}, {Bruel}, {Burnett},
  {Buson}, {Caliandro}, {Cameron}, {Caraveo}, {Casandjian}, {Cecchi}, {{\c
  C}elik}, {Chekhtman}, {Cheung}, {Chiang}, {Ciprini}, {Claus}, {Cohen-Tanugi},
  {Cominsky}, {Conrad}, {Cutini}, {Dermer}, {de Angelis}, {de Palma}, {Digel},
  {Dormody}, {Silva}, {Drell}, {Dubois}, {Dumora}, {Farnier}, {Favuzzi},
  {Fegan}, {Focke}, {Fortin}, {Frailis}, {Fukazawa}, {Funk}, {Fusco},
  {Gargano}, {Gasparrini}, {Gehrels}, {Germani}, {Giavitto}, {Giebels},
  {Giglietto}, {Giordano}, {Glanzman}, {Godfrey}, {Grenier}, {Grondin},
  {Grove}, {Guillemot}, {Guiriec}, {Hanabata}, {Harding}, {Hayashida}, {Hays},
  {Hughes}, {Jackson}, {J{\'o}hannesson}, {Johnson}, {Johnson}, {Johnson},
  {Kamae}, {Katagiri}, {Kataoka}, {Katsuta}, {Kawai}, {Kerr}, {Kn{\"o}dlseder},
  {Kocian}, {Kuss}, {Lande}, {Latronico}, {Lemoine-Goumard}, {Longo},
  {Loparco}, {Lott}, {Lovellette}, {Lubrano}, {Makeev}, {Mazziotta}, {McEnery},
  {Meurer}, {Michelson}, {Mitthumsiri}, {Mizuno}, {Moiseev}, {Monte},
  {Monzani}, {Morselli}, {Moskalenko}, {Murgia}, {Nakamori}, {Nolan}, {Norris},
  {Nuss}, {Ohsugi}, {Okumura}, {Omodei}, {Orlando}, {Ormes}, {Paneque},
  {Parent}, {Pelassa}, {Pepe}, {Pesce-Rollins}, {Piron}, {Porter}, {Rain{\`o}},
  {Rando}, {Razzano}, {Reimer}, {Reimer}, {Reposeur}, {Ritz}, {Rodriguez},
  {Romani}, {Roth}, {Ryde}, {Sadrozinski}, {Sanchez}, {Sander}, {Saz
  Parkinson}, {Scargle}, {Schalk}, {Sgr{\`o}}, {Siskind}, {Smith}, {Smith},
  {Spandre}, {Spinelli}, {Strickman}, {Suson}, {Tajima}, {Takahashi},
  {Takahashi}, {Tanaka}, {Thayer}, {Thayer}, {Thompson}, {Tibaldo}, {Tibolla},
  {Torres}, {Tosti}, {Tramacere}, {Uchiyama}, {Usher}, {Vasileiou}, {Venter},
  {Vilchez}, {Vitale}, {Waite}, {Wang}, {Winer}, {Wood}, {Yamazaki}, {Ylinen},
  \& {Ziegler}}]{Abdo09_W51C}
---. 2009{\natexlab{b}}, \apjl, 706, L1

\bibitem[{{Abdo} {et~al.}(2009{\natexlab{c}}){Abdo}, {Ackermann}, {Ajello},
  {Atwood}, {Axelsson}, {Baldini}, {Ballet}, {Barbiellini}, {Bastieri},
  {Baughman}, {Bechtol}, {Bellazzini}, {Berenji}, {Bloom}, {Bonamente},
  {Borgland}, {Bregeon}, {Brez}, {Brigida}, {Bruel}, {Burnett}, {Caliandro},
  {Cameron}, {Caraveo}, {Carlson}, {Casandjian}, {Cecchi}, {{\c C}elik},
  {Chekhtman}, {Cheung}, {Ciprini}, {Claus}, {Cohen-Tanugi}, {Conrad},
  {Cutini}, {Dermer}, {de Angelis}, {de Palma}, {Digel}, {Silva}, {Drell},
  {Dubois}, {Dumora}, {Farnier}, {Favuzzi}, {Fegan}, {Focke}, {Frailis},
  {Fukazawa}, {Funk}, {Fusco}, {Gargano}, {Gasparrini}, {Gehrels}, {Germani},
  {Giebels}, {Giglietto}, {Giordano}, {Glanzman}, {Godfrey}, {Grenier},
  {Grondin}, {Grove}, {Guillemot}, {Guiriec}, {Hanabata}, {Harding},
  {Hayashida}, {Hays}, {Hughes}, {J{\'o}hannesson}, {Johnson}, {Johnson},
  {Johnson}, {Kamae}, {Katagiri}, {Kawai}, {Kerr}, {Kn{\"o}dlseder}, {Kocian},
  {Kuehn}, {Kuss}, {Lande}, {Latronico}, {Lemoine-Goumard}, {Longo}, {Loparco},
  {Lott}, {Lovellette}, {Lubrano}, {Makeev}, {Mazziotta}, {McEnery}, {Meurer},
  {Michelson}, {Mitthumsiri}, {Mizuno}, {Moiseev}, {Monte}, {Monzani},
  {Morselli}, {Moskalenko}, {Murgia}, {Nolan}, {Norris}, {Nuss}, {Ohsugi},
  {Okumura}, {Omodei}, {Orlando}, {Ormes}, {Ozaki}, {Paneque}, {Panetta},
  {Parent}, {Pepe}, {Pesce-Rollins}, {Piron}, {Pohl}, {Porter}, {Rain{\`o}},
  {Rando}, {Razzano}, {Reimer}, {Reimer}, {Reposeur}, {Ritz}, {Rochester},
  {Rodriguez}, {Ryde}, {Sadrozinski}, {Sanchez}, {Sander}, {Saz Parkinson},
  {Schalk}, {Sellerholm}, {Sgr{\`o}}, {Smith}, {Smith}, {Spandre}, {Spinelli},
  {Starck}, {Stecker}, {Strickman}, {Strong}, {Suson}, {Tajima}, {Takahashi},
  {Takahashi}, {Tanaka}, {Thayer}, {Thayer}, {Thompson}, {Tibaldo}, {Torres},
  {Tosti}, {Tramacere}, {Uchiyama}, {Usher}, {Vasileiou}, {Vilchez}, {Vitale},
  {Waite}, {Wang}, {Winer}, {Wood}, {Ylinen}, \& {Ziegler}}]{Abdo09_CRint}
---. 2009{\natexlab{c}}, \apj, 703, 1249

\bibitem[{{Abdo} {et~al.}(2010){Abdo}, {Ackermann}, {Ajello}, {Allafort},
  {Baldini}, {Ballet}, {Barbiellini}, {Baring}, {Bastieri}, {Baughman},
  {Bechtol}, {Bellazzini}, {Berenji}, {Blandford}, {Bloom}, {Bonamente},
  {Borgland}, {Bregeon}, {Brez}, {Brigida}, {Bruel}, {Buehler}, {Burnett},
  {Busetto}, {Caliandro}, {Cameron}, {Caraveo}, {Casandjian}, {Cecchi}, {{\c
  C}elik}, {Charles}, {Chaty}, {Chekhtman}, {Cheung}, {Chiang}, {Cillis},
  {Ciprini}, {Claus}, {Cohen-Tanugi}, {Conrad}, {Corbel}, {de Palma}, {Digel},
  {Dormody}, {Silva}, {Drell}, {Dubois}, {Dumora}, {Edmonds}, {Farnier},
  {Favuzzi}, {Fegan}, {Ferrara}, {Focke}, {Fortin}, {Frailis}, {Fukazawa},
  {Funk}, {Fusco}, {Gargano}, {Gasparrini}, {Gehrels}, {Germani}, {Giavitto},
  {Giglietto}, {Giordano}, {Glanzman}, {Godfrey}, {Grenier}, {Grondin},
  {Grove}, {Guillemot}, {Guiriec}, {Hanabata}, {Hays}, {Harding}, {Hayashida},
  {Horan}, {Hughes}, {Jackson}, {Johnson}, {Johnson}, {Johnson}, {Kamae},
  {Katagiri}, {Kataoka}, {Kawai}, {Kerr}, {Kn{\"o}dlseder}, {Kuss}, {Lande},
  {Latronico}, {Lemoine-Goumard}, {Longo}, {Loparco}, {Lott}, {Lovellette},
  {Lubrano}, {Makeev}, {Mazziotta}, {Meurer}, {Michelson}, {Mitthumsiri},
  {Mizuno}, {Monte}, {Monzani}, {Morselli}, {Moskalenko}, {Murgia}, {Nakamori},
  {Nolan}, {Norris}, {Nuss}, {Ohsugi}, {Okumura}, {Omodei}, {Orlando}, {Ormes},
  {Paneque}, {Panetta}, {Pelassa}, {Pepe}, {Pesce-Rollins}, {Piron}, {Pohl},
  {Porter}, {Rain{\`o}}, {Rando}, {Reimer}, {Reimer}, {Reposeur}, {Ritz},
  {Rodriguez}, {Romani}, {Roth}, {Sadrozinski}, {Sander}, {Saz Parkinson},
  {Scargle}, {Sgr{\`o}}, {Siskind}, {Smith}, {Smith}, {Spinelli}, {Strickman},
  {Suson}, {Tajima}, {Takahashi}, {Tanaka}, {Thayer}, {Thayer}, {Thompson},
  {Thorsett}, {Tibaldo}, {Tibolla}, {Torres}, {Tosti}, {Tramacere}, {Uchiyama},
  {Usher}, {Van Etten}, {Vasileiou}, {Venter}, {Vilchez}, {Vitale}, {Waite},
  {Wang}, {Winer}, {Wood}, {Yamazaki}, {Ylinen}, \& {Ziegler}}]{Abdo10_CasA}
---. 2010, \apjl, 710, L92

\bibitem[{{Acciari} {et~al.}(2010){Acciari}, {Aliu}, {Arlen}, {Aune},
  {Bautista}, {Beilicke}, {Benbow}, {Boltuch}, {Bradbury}, {Buckley}, {Bugaev},
  {Butt}, {Byrum}, {Cannon}, {Cesarini}, {Chow}, {Ciupik}, {Cogan}, {Cui},
  {Dickherber}, {Duke}, {Ergin}, {Fegan}, {Finley}, {Finnegan}, {Fortin},
  {Fortson}, {Furniss}, {Galante}, {Gall}, {Gillanders}, {Grube}, {Guenette},
  {Gyuk}, {Hanna}, {Holder}, {Huang}, {Hui}, {Humensky}, {Kaaret}, {Karlsson},
  {Kertzman}, {Kieda}, {Konopelko}, {Krawczynski}, {Krennrich}, {Lang},
  {LeBohec}, {Maier}, {McArthur}, {McCann}, {McCutcheon}, {Millis}, {Moriarty},
  {Ong}, {Pandel}, {Perkins}, {Pohl}, {Quinn}, {Ragan}, {Reynolds}, {Roache},
  {Rose}, {Schroedter}, {Sembroski}, {Smith}, {Smith}, {Steele}, {Swordy},
  {Theiling}, {Thibadeau}, {Varlotta}, {Vassiliev}, {Vincent}, {Wagner},
  {Wakely}, {Ward}, {Weekes}, {Weinstein}, {Weisgarber}, {Wissel}, {Wood}, \&
  {VERITAS Collaboration}}]{Acciari11_CasA}
{Acciari}, V.~A., {et~al.} 2010, \apj, 714, 163

\bibitem[{{Acciari} {et~al.}(2011){Acciari}, {Aliu}, {Arlen}, {Aune},
  {Beilicke}, {Benbow}, {Bradbury}, {Buckley}, {Bugaev}, {Byrum}, {Cannon},
  {Cesarini}, {Ciupik}, {Collins-Hughes}, {Cui}, {Dickherber}, {Duke},
  {Errando}, {Finley}, {Finnegan}, {Fortson}, {Furniss}, {Galante}, {Gall},
  {Gillanders}, {Godambe}, {Griffin}, {Grube}, {Guenette}, {Gyuk}, {Hanna},
  {Holder}, {Hughes}, {Hui}, {Humensky}, {Kaaret}, {Karlsson}, {Kertzman},
  {Kieda}, {Krawczynski}, {Krennrich}, {Lang}, {LeBohec}, {Madhavan}, {Maier},
  {Majumdar}, {McArthur}, {McCann}, {Moriarty}, {Mukherjee}, {Ong}, {Orr},
  {Otte}, {Pandel}, {Park}, {Perkins}, {Pohl}, {Quinn}, {Ragan}, {Reyes},
  {Reynolds}, {Roache}, {Rose}, {Saxon}, {Schroedter}, {Sembroski}, {Senturk},
  {Slane}, {Smith}, {Te{\v s}i{\'c}}, {Theiling}, {Thibadeau}, {Tsurusaki},
  {Varlotta}, {Vassiliev}, {Vincent}, {Vivier}, {Wakely}, {Ward}, {Weekes},
  {Weinstein}, {Weisgarber}, {Williams}, {Wood}, \& {Zitzer}}]{Acciari11_Tycho}
---. 2011, \apjl, 730, L20+

\bibitem[{{Acero} {et~al.}(2010){Acero}, {Aharonian}, {Akhperjanian}, {Anton},
  {Barres de Almeida}, {Bazer-Bachi}, {Becherini}, {Behera}, {Beilicke},
  {Bernl{\"o}hr}, {Bochow}, {Boisson}, {Bolmont}, {Borrel}, {Brucker}, {Brun},
  {Brun}, {B{\"u}hler}, {Bulik}, {B{\"u}sching}, {Boutelier}, {Chadwick},
  {Charbonnier}, {Chaves}, {Cheesebrough}, {Conrad}, {Chounet}, {Clapson},
  {Coignet}, {Dalton}, {Daniel}, {Davids}, {Degrange}, {Deil}, {Dickinson},
  {Djannati-Ata{\"i}}, {Domainko}, {O'C.~Drury}, {Dubois}, {Dubus}, {Dyks},
  {Dyrda}, {Egberts}, {Eger}, {Espigat}, {Fallon}, {Farnier}, {Fegan},
  {Feinstein}, {Fiasson}, {F{\"o}rster}, {Fontaine}, {F{\"u}{\ss}ling},
  {Gabici}, {Gallant}, {G{\'e}rard}, {Gerbig}, {Giebels}, {Glicenstein},
  {Gl{\"u}ck}, {Goret}, {G{\"o}ring}, {Hauser}, {Hauser}, {Heinz},
  {Heinzelmann}, {Henri}, {Hermann}, {Hinton}, {Hoffmann}, {Hofmann},
  {Hofverberg}, {Holleran}, {Hoppe}, {Horns}, {Jacholkowska}, {de Jager},
  {Jahn}, {Jung}, {Katarzy{\'n}ski}, {Katz}, {Kaufmann}, {Kerschhaggl},
  {Khangulyan}, {Kh{\'e}lifi}, {Keogh}, {Klochkov}, {Klu{\'z}niak}, {Kneiske},
  {Komin}, {Kosack}, {Kossakowski}, {Lamanna}, {Lemoine-Goumard}, {Lenain},
  {Lohse}, {Marandon}, {Marcowith}, {Masbou}, {Maurin}, {McComb}, {Medina},
  {M{\'e}hault}, {Moderski}, {Moulin}, {Naumann-Godo}, {de Naurois}, {Nedbal},
  {Nekrassov}, {Nicholas}, {Niemiec}, {Nolan}, {Ohm}, {Olive}, {de O{\~n}a
  Wilhelmi}, {Orford}, {Ostrowski}, {Panter}, {Paz Arribas}, {Pedaletti},
  {Pelletier}, {Petrucci}, {Pita}, {P{\"u}hlhofer}, {Punch}, {Quirrenbach},
  {Raubenheimer}, {Raue}, {Rayner}, {Reimer}, {Renaud}, {de Los Reyes},
  {Rieger}, {Ripken}, {Rob}, {Rosier-Lees}, {Rowell}, {Rudak}, {Rulten},
  {Ruppel}, {Ryde}, {Sahakian}, {Santangelo}, {Schlickeiser}, {Sch{\"o}ck},
  {Sch{\"o}nwald}, {Schwanke}, {Schwarzburg}, {Schwemmer}, {Shalchi}, {Sushch},
  {Sikora}, {Skilton}, {Sol}, {Stawarz}, {Steenkamp}, {Stegmann}, {Stinzing},
  {Superina}, {Szostek}, {Tam}, {Tavernet}, {Terrier}, {Tibolla}, {Tluczykont},
  {van Eldik}, {Vasileiadis}, {Venter}, {Venter}, {Vialle}, {Vincent}, {Vink},
  {Vivier}, {V{\"o}lk}, {Volpe}, {Vorobiov}, {Wagner}, {Ward}, {Zdziarski},
  {Zech}, \& {HESS Collaboration}}]{Acero10}
{Acero}, F., {et~al.} 2010, \aap, 516, A62+

\bibitem[{{Ando} \& {Pavlidou}(2009)}]{Ando09_EGB}
{Ando}, S., \& {Pavlidou}, V. 2009, \mnras, 400, 2122

\bibitem[{{Baade} \& {Zwicky}(1934)}]{Baade34}
{Baade}, W., \& {Zwicky}, F. 1934, Proceedings of the National Academy of
  Science, 20, 259

\bibitem[{{Bailey} {et~al.}(2009){Bailey}, {Bernstein}, {Cinabro}, {Kessler},
  \& {Kuhlma}}]{LSSTsb_Ia}
{Bailey}, S., {Bernstein}, J.~P., {Cinabro}, D., {Kessler}, R., \& {Kuhlma}, S.
  2009, ArXiv e-prints: 0912.0201

\bibitem[{{Bazin} {et~al.}(2009){Bazin}, {Palanque-Delabrouille}, {Rich},
  {Ruhlmann-Kleider}, {Aubourg}, {Le Guillou}, {Astier}, {Balland}, {Basa},
  {Carlberg}, {Conley}, {Fouchez}, {Guy}, {Hardin}, {Hook}, {Howell}, {Pain},
  {Perrett}, {Pritchet}, {Regnault}, {Sullivan}, {Antilogus}, {Arsenijevic},
  {Baumont}, {Fabbro}, {Le Du}, {Lidman}, {Mouchet}, {Mour{\~a}o}, \&
  {Walker}}]{Bazin09}
{Bazin}, G., {et~al.} 2009, \aap, 499, 653

\bibitem[{{Berezhko} \& {Ellison}(1999)}]{Berezhko99}
{Berezhko}, E.~G., \& {Ellison}, D.~C. 1999, \apj, 526, 385

\bibitem[{{Bregman} {et~al.}(1992){Bregman}, {Hogg}, \& {Roberts}}]{Bregman92}
{Bregman}, J.~N., {Hogg}, D.~E., \& {Roberts}, M.~S. 1992, \apj, 387, 484

\bibitem[{{Canizares} {et~al.}(1987){Canizares}, {Fabbiano}, \&
  {Trinchieri}}]{Canizares87}
{Canizares}, C.~R., {Fabbiano}, G., \& {Trinchieri}, G. 1987, \apj, 312, 503

\bibitem[{{Capelo} {et~al.}(2010){Capelo}, {Natarajan}, \& {Coppi}}]{Capelo10}
{Capelo}, P.~R., {Natarajan}, P., \& {Coppi}, P.~S. 2010, \mnras, 407, 1148

\bibitem[{{Dahlen} {et~al.}(2008){Dahlen}, {Strolger}, \&
  {Riess}}]{Dahlen08_Ia}
{Dahlen}, T., {Strolger}, L.-G., \& {Riess}, A.~G. 2008, \apj, 681, 462

\bibitem[{{Dar} \& {Shaviv}(1995)}]{Dar95}
{Dar}, A., \& {Shaviv}, N.~J. 1995, Physical Review Letters, 75, 3052

\bibitem[{{David} {et~al.}(2006){David}, {Jones}, {Forman}, {Vargas}, \&
  {Nulsen}}]{David06}
{David}, L.~P., {Jones}, C., {Forman}, W., {Vargas}, I.~M., \& {Nulsen}, P.
  2006, \apj, 653, 207

\bibitem[{{Dermer}(2007{\natexlab{a}})}]{Dermer07AGN}
{Dermer}, C.~D. 2007{\natexlab{a}}, \apj, 659, 958

\bibitem[{{Dermer}(2007{\natexlab{b}})}]{Dermer07}
{Dermer}, C.~D. 2007{\natexlab{b}}, in American Institute of Physics Conference
  Series, Vol. 921, The First GLAST Symposium, ed. {S.~Ritz, P.~Michelson, \&
  C.~A.~Meegan}, 122--126

\bibitem[{{Dilday} {et~al.}(2010{\natexlab{a}}){Dilday}, {Bassett}, {Becker},
  {Bender}, {Castander}, {Cinabro}, {Frieman}, {Galbany}, {Garnavich},
  {Goobar}, {Hopp}, {Ihara}, {Jha}, {Kessler}, {Lampeitl}, {Marriner},
  {Miquel}, {Moll{\'a}}, {Nichol}, {Nordin}, {Riess}, {Sako}, {Schneider},
  {Smith}, {Sollerman}, {Wheeler}, {{\"O}stman}, {Bizyaev}, {Brewington},
  {Malanushenko}, {Malanushenko}, {Oravetz}, {Pan}, {Simmons}, \&
  {Snedden}}]{Dilday10_Iacluster}
{Dilday}, B., {et~al.} 2010{\natexlab{a}}, \apj, 715, 1021

\bibitem[{{Dilday} {et~al.}(2010{\natexlab{b}}){Dilday}, {Smith}, {Bassett},
  {Becker}, {Bender}, {Castander}, {Cinabro}, {Filippenko}, {Frieman},
  {Galbany}, {Garnavich}, {Goobar}, {Hopp}, {Ihara}, {Jha}, {Kessler},
  {Lampeitl}, {Marriner}, {Miquel}, {Moll{\'a}}, {Nichol}, {Nordin}, {Riess},
  {Sako}, {Schneider}, {Sollerman}, {Wheeler}, {{\"O}stman}, {Bizyaev},
  {Brewington}, {Malanushenko}, {Malanushenko}, {Oravetz}, {Pan}, {Simmons}, \&
  {Snedden}}]{Dilday10}
---. 2010{\natexlab{b}}, \apj, 713, 1026

\bibitem[{{Dorfi} \& {Voelk}(1996)}]{Dorfi96}
{Dorfi}, E.~A., \& {Voelk}, H.~J. 1996, \aap, 307, 715

\bibitem[{{Ellison} {et~al.}(1997){Ellison}, {Drury}, \& {Meyer}}]{Ellison97}
{Ellison}, D.~C., {Drury}, L.~O., \& {Meyer}, J.-P. 1997, \apj, 487, 197

\bibitem[{{Ellison} {et~al.}(2007){Ellison}, {Patnaude}, {Slane}, {Blasi}, \&
  {Gabici}}]{Ellison07}
{Ellison}, D.~C., {Patnaude}, D.~J., {Slane}, P., {Blasi}, P., \& {Gabici}, S.
  2007, \apj, 661, 879

\bibitem[{{Ellison} {et~al.}(2012){Ellison}, {Slane}, {Patnaude}, \&
  {Bykov}}]{Ellison12}
{Ellison}, D.~C., {Slane}, P., {Patnaude}, D.~J., \& {Bykov}, A.~M. 2012, \apj,
  744, 39

\bibitem[{{Fichtel} {et~al.}(1977){Fichtel}, {Hartman}, {Kniffen}, {Thompson},
  {Ogelman}, {Ozel}, \& {Tumer}}]{Fichtel77}
{Fichtel}, C.~E., {Hartman}, R.~C., {Kniffen}, D.~A., {Thompson}, D.~J.,
  {Ogelman}, H.~B., {Ozel}, M.~E., \& {Tumer}, T. 1977, \apjl, 217, L9

\bibitem[{{Fichtel} {et~al.}(1978){Fichtel}, {Simpson}, \&
  {Thompson}}]{Fichtel78}
{Fichtel}, C.~E., {Simpson}, G.~A., \& {Thompson}, D.~J. 1978, \apj, 222, 833

\bibitem[{{Fields} {et~al.}(2001){Fields}, {Olive}, {Cass{\'e}}, \&
  {Vangioni-Flam}}]{Fields01}
{Fields}, B.~D., {Olive}, K.~A., {Cass{\'e}}, M., \& {Vangioni-Flam}, E. 2001,
  \aap, 370, 623

\bibitem[{{Fields} {et~al.}(2010){Fields}, {Pavlidou}, \&
  {Prodanovi{\'c}}}]{Fields10}
{Fields}, B.~D., {Pavlidou}, V., \& {Prodanovi{\'c}}, T. 2010, \apjl, 722, L199

\bibitem[{{Filippenko}(2001)}]{Filippenko01}
{Filippenko}, A.~V. 2001, in American Institute of Physics Conference Series,
  Vol. 565, Young Supernova Remnants, ed. {S.~S.~Holt \& U.~Hwang}, 40--58

\bibitem[{{Forman} {et~al.}(1985){Forman}, {Jones}, \& {Tucker}}]{Forman85}
{Forman}, W., {Jones}, C., \& {Tucker}, W. 1985, \apj, 293, 102

\bibitem[{{Fukazawa} {et~al.}(2006){Fukazawa}, {Botoya-Nonesa}, {Pu}, {Ohto},
  \& {Kawano}}]{Fukazawa06}
{Fukazawa}, Y., {Botoya-Nonesa}, J.~G., {Pu}, J., {Ohto}, A., \& {Kawano}, N.
  2006, \apj, 636, 698

\bibitem[{{Gal-Yam} \& {Maoz}(2004)}]{gal-yam}
{Gal-Yam}, A., \& {Maoz}, D. 2004, \mnras, 347, 942

\bibitem[{{Gal-Yam} {et~al.}(2002){Gal-Yam}, {Maoz}, \& {Sharon}}]{Gal-Yam02}
{Gal-Yam}, A., {Maoz}, D., \& {Sharon}, K. 2002, \mnras, 332, 37

\bibitem[{{Ginzburg} \& {Syrovatskii}(1964)}]{Ginzburg64}
{Ginzburg}, V.~L., \& {Syrovatskii}, S.~I. 1964, {The Origin of Cosmic Rays}
  (New York: Macmillan)

\bibitem[{{Gnedin} \& {Ostriker}(1992)}]{Gnedin92}
{Gnedin}, N.~I., \& {Ostriker}, J.~P. 1992, \apj, 400, 1

\bibitem[{{Graham} {et~al.}(2008){Graham}, {Pritchet}, {Sullivan}, {Gwyn},
  {Neill}, {Hsiao}, {Astier}, {Balam}, {Balland}, {Basa}, {Carlberg}, {Conley},
  {Fouchez}, {Guy}, {Hardin}, {Hook}, {Howell}, {Pain}, {Perrett}, {Regnault},
  {Baumont}, {LeDu}, {Lidman}, {Perlmutter}, {Ripoche}, {Suzuki}, {Walker}, \&
  {Zhang}}]{Graham08}
{Graham}, M.~L., {et~al.} 2008, \aj, 135, 1343

\bibitem[{{Graur} {et~al.}(2011){Graur}, {Poznanski}, {Maoz}, {Yasuda},
  {Totani}, {Fukugita}, {Filippenko}, {Foley}, {Silverman}, {Gal-Yam},
  {Horesh}, \& {Jannuzi}}]{Graur11}
{Graur}, O., {et~al.} 2011, \mnras, 417, 916

\bibitem[{{Hein} \& {Spanier}(2008)}]{Hein08}
{Hein}, T., \& {Spanier}, F. 2008, \aap, 481, 1

\bibitem[{{Helder} {et~al.}(2010){Helder}, {Kosenko}, \& {Vink}}]{Helder10}
{Helder}, E.~A., {Kosenko}, D., \& {Vink}, J. 2010, \apjl, 719, L140

\bibitem[{{Hopkins}(2004)}]{Hopkins04}
{Hopkins}, A.~M. 2004, \apj, 615, 209

\bibitem[{{Hopkins} \& {Beacom}(2006)}]{Hopkins06}
{Hopkins}, A.~M., \& {Beacom}, J.~F. 2006, \apj, 651, 142

\bibitem[{{Horiuchi} \& {Beacom}(2010)}]{Horiuchi10}
{Horiuchi}, S., \& {Beacom}, J.~F. 2010, \apj, 723, 329

\bibitem[{{Horiuchi} {et~al.}(2009){Horiuchi}, {Beacom}, \&
  {Dwek}}]{Horiuchi09}
{Horiuchi}, S., {Beacom}, J.~F., \& {Dwek}, E. 2009, \prd, 79, 083013

\bibitem[{{Humphrey} \& {Buote}(2010)}]{Humphrey10}
{Humphrey}, P.~J., \& {Buote}, D.~A. 2010, \mnras, 403, 2143

\bibitem[{{Humphrey} {et~al.}(2011){Humphrey}, {Buote}, {Canizares}, {Fabian},
  \& {Miller}}]{Humphrey11}
{Humphrey}, P.~J., {Buote}, D.~A., {Canizares}, C.~R., {Fabian}, A.~C., \&
  {Miller}, J.~M. 2011, \apj, 729, 53

\bibitem[{{Hunter} {et~al.}(1997){Hunter}, {Bertsch}, {Catelli}, {Dame},
  {Digel}, {Dingus}, {Esposito}, {Fichtel}, {Hartman}, {Kanbach}, {Kniffen},
  {Lin}, {Mayer-Hasselwander}, {Michelson}, {von Montigny}, {Mukherjee},
  {Nolan}, {Schneid}, {Sreekumar}, {Thaddeus}, \& {Thompson}}]{Hunter97}
{Hunter}, S.~D., {et~al.} 1997, \apj, 481, 205

\bibitem[{{Iben} \& {Tutukov}(1984)}]{Iben84}
{Iben}, Jr., I., \& {Tutukov}, A.~V. 1984, \apjs, 54, 335

\bibitem[{{Inoue} \& {Totani}(2009)}]{Inoue09}
{Inoue}, Y., \& {Totani}, T. 2009, \apj, 702, 523

\bibitem[{{Ivezic} {et~al.}(2008){Ivezic}, {Tyson}, {Acosta}, {Allsman},
  {Anderson}, {Andrew}, {Angel}, {Axelrod}, {Barr}, {Becker}, {Becla},
  {Beldica}, {Blandford}, {Bloom}, {Borne}, {Brandt}, {Brown}, {Bullock},
  {Burke}, {Chandrasekharan}, {Chesley}, {Claver}, {Connolly}, {Cook},
  {Cooray}, {Covey}, {Cribbs}, {Cutri}, {Daues}, {Delgado}, {Ferguson},
  {Gawiser}, {Geary}, {Gee}, {Geha}, {Gibson}, {Gilmore}, {Gressler}, {Hogan},
  {Huffer}, {Jacoby}, {Jain}, {Jernigan}, {Jones}, {Juric}, {Kahn}, {Kalirai},
  {Kantor}, {Kessler}, {Kirkby}, {Knox}, {Krabbendam}, {Krughoff}, {Kulkarni},
  {Lambert}, {Levine}, {Liang}, {Lim}, {Lupton}, {Marshall}, {Marshall}, {May},
  {Miller}, {Mills}, {Monet}, {Neill}, {Nordby}, {O'Connor}, {Oliver},
  {Olivier}, {Olsen}, {Owen}, {Peterson}, {Petry}, {Pierfederici},
  {Pietrowicz}, {Pike}, {Pinto}, {Plante}, {Radeka}, {Rasmussen}, {Ridgway},
  {Rosing}, {Saha}, {Schalk}, {Schindler}, {Schneider}, {Schumacher}, {Sebag},
  {Seppala}, {Shipsey}, {Silvestri}, {Smith}, {Smith}, {Strauss}, {Stubbs},
  {Sweeney}, {Szalay}, {Thaler}, {Vanden Berk}, {Walkowicz}, {Warner},
  {Willman}, {Wittman}, {Wolff}, {Wood-Vasey}, {Yoachim}, {Zhan}, \& {for the
  LSST Collaboration}}]{LSST_overview}
{Ivezic}, Z., {et~al.} 2008, ArXiv e-prints: 0805.2366

\bibitem[{{Jiang} \& {Kochanek}(2007)}]{Jiang07}
{Jiang}, G., \& {Kochanek}, C.~S. 2007, \apj, 671, 1568

\bibitem[{{Komatsu} {et~al.}(2011){Komatsu}, {Smith}, {Dunkley}, {Bennett},
  {Gold}, {Hinshaw}, {Jarosik}, {Larson}, {Nolta}, {Page}, {Spergel},
  {Halpern}, {Hill}, {Kogut}, {Limon}, {Meyer}, {Odegard}, {Tucker}, {Weiland},
  {Wollack}, \& {Wright}}]{wmap7}
{Komatsu}, E., {et~al.} 2011, \apjs, 192, 18

\bibitem[{{Kuznetsova} {et~al.}(2008){Kuznetsova}, {Barbary}, {Connolly},
  {Kim}, {Pain}, {Roe}, {Aldering}, {Amanullah}, {Dawson}, {Doi}, {Fadeyev},
  {Fruchter}, {Gibbons}, {Goldhaber}, {Goobar}, {Gude}, {Knop}, {Kowalski},
  {Lidman}, {Morokuma}, {Meyers}, {Perlmutter}, {Rubin}, {Schlegel},
  {Spadafora}, {Stanishev}, {Strovink}, {Suzuki}, {Wang}, {Yasuda}, \&
  {Supernova Cosmology Project}}]{Kuznetsova08}
{Kuznetsova}, N., {et~al.} 2008, \apj, 673, 981

\bibitem[{{Lacki} {et~al.}(2011){Lacki}, {Thompson}, {Quataert}, {Loeb}, \&
  {Waxman}}]{Lacki10}
{Lacki}, B.~C., {Thompson}, T.~A., {Quataert}, E., {Loeb}, A., \& {Waxman}, E.
  2011, \apj, 734, 107

\bibitem[{{Lenain} \& {Walter}(2011)}]{Lenain11}
{Lenain}, J.-P., \& {Walter}, R. 2011, \aap, 535, A19

\bibitem[{{Lien} \& {Fields}(2009)}]{lf}
{Lien}, A., \& {Fields}, B.~D. 2009, \jcap, 1, 47

\bibitem[{{Makiya} {et~al.}(2011){Makiya}, {Totani}, \& {Kobayashi}}]{Makiya11}
{Makiya}, R., {Totani}, T., \& {Kobayashi}, M.~A.~R. 2011, \apj, 728, 158

\bibitem[{{Mannucci} {et~al.}(2006){Mannucci}, {Della Valle}, \&
  {Panagia}}]{Mannucci06}
{Mannucci}, F., {Della Valle}, M., \& {Panagia}, N. 2006, \mnras, 370, 773

\bibitem[{{Mannucci} {et~al.}(2005){Mannucci}, {Della Valle}, {Panagia},
  {Cappellaro}, {Cresci}, {Maiolino}, {Petrosian}, \& {Turatto}}]{Mannucci05}
{Mannucci}, F., {Della Valle}, M., {Panagia}, N., {Cappellaro}, E., {Cresci},
  G., {Maiolino}, R., {Petrosian}, A., \& {Turatto}, M. 2005, \aap, 433, 807

\bibitem[{{Mannucci} {et~al.}(2008){Mannucci}, {Maoz}, {Sharon}, {Botticella},
  {Della Valle}, {Gal-Yam}, \& {Panagia}}]{Mannucci08}
{Mannucci}, F., {Maoz}, D., {Sharon}, K., {Botticella}, M.~T., {Della Valle},
  M., {Gal-Yam}, A., \& {Panagia}, N. 2008, \mnras, 383, 1121

\bibitem[{{Maoz} {et~al.}(2011){Maoz}, {Mannucci}, {Li}, {Filippenko}, {Valle},
  \& {Panagia}}]{Maoz11}
{Maoz}, D., {Mannucci}, F., {Li}, W., {Filippenko}, A.~V., {Valle}, M.~D., \&
  {Panagia}, N. 2011, \mnras, 412, 1508

\bibitem[{{Mukherjee} \& {Chiang}(1999)}]{Mukherjee99}
{Mukherjee}, R., \& {Chiang}, J. 1999, Astroparticle Physics, 11, 213

\bibitem[{{Nakamura} {et~al.}(2004){Nakamura}, {Fukugita}, {Brinkmann}, \&
  {Schneider}}]{Nakamura04}
{Nakamura}, O., {Fukugita}, M., {Brinkmann}, J., \& {Schneider}, D.~P. 2004,
  \aj, 127, 2511

\bibitem[{{Nomoto} {et~al.}(1984){Nomoto}, {Thielemann}, \& {Yokoi}}]{Nomoto84}
{Nomoto}, K., {Thielemann}, F.-K., \& {Yokoi}, K. 1984, \apj, 286, 644

\bibitem[{{Padovani} {et~al.}(1993){Padovani}, {Ghisellini}, {Fabian}, \&
  {Celotti}}]{Padovani93}
{Padovani}, P., {Ghisellini}, G., {Fabian}, A.~C., \& {Celotti}, A. 1993,
  \mnras, 260, L21

\bibitem[{{Page} \& {Hawking}(1976)}]{Page76}
{Page}, D.~N., \& {Hawking}, S.~W. 1976, \apj, 206, 1

\bibitem[{{Pannella} {et~al.}(2009){Pannella}, {Gabasch}, {Goranova}, {Drory},
  {Hopp}, {Noll}, {Saglia}, {Strazzullo}, \& {Bender}}]{Pannella09}
{Pannella}, M., {et~al.} 2009, \apj, 701, 787

\bibitem[{{Pavlidou} \& {Fields}(2001)}]{Pavlidou01}
{Pavlidou}, V., \& {Fields}, B.~D. 2001, \apj, 558, 63

\bibitem[{{Pavlidou} \& {Fields}(2002)}]{Pavlidou02}
---. 2002, \apjl, 575, L5

\bibitem[{{Pavlidou} \& {Venters}(2008)}]{Pavlidou08}
{Pavlidou}, V., \& {Venters}, T.~M. 2008, \apj, 673, 114

\bibitem[{{Prodanovi{\'c}} \& {Fields}(2006)}]{Prodanovic06}
{Prodanovi{\'c}}, T., \& {Fields}, B.~D. 2006, \apjl, 645, L125

\bibitem[{{Reynolds} \& {Ellison}(1992)}]{Reynolds92}
{Reynolds}, S.~P., \& {Ellison}, D.~C. 1992, \apjl, 399, L75

\bibitem[{{Robitaille} \& {Whitney}(2010)}]{Robitaille10}
{Robitaille}, T.~P., \& {Whitney}, B.~A. 2010, \apjl, 710, L11

\bibitem[{{Rudaz} \& {Stecker}(1991)}]{Rudaz91}
{Rudaz}, S., \& {Stecker}, F.~W. 1991, \apj, 368, 406

\bibitem[{{Salamon} \& {Stecker}(1998)}]{Salamon98}
{Salamon}, M.~H., \& {Stecker}, F.~W. 1998, \apj, 493, 547

\bibitem[{{Scannapieco} \& {Bildsten}(2005)}]{Scannapieco05}
{Scannapieco}, E., \& {Bildsten}, L. 2005, \apjl, 629, L85

\bibitem[{{Schlickeiser}(1989)}]{Schlickeiser89}
{Schlickeiser}, R. 1989, \apj, 336, 243

\bibitem[{{Sharon} {et~al.}(2007){Sharon}, {Gal-Yam}, {Maoz}, {Filippenko}, \&
  {Guhathakurta}}]{Sharon07}
{Sharon}, K., {Gal-Yam}, A., {Maoz}, D., {Filippenko}, A.~V., \&
  {Guhathakurta}, P. 2007, \apj, 660, 1165

\bibitem[{{Sharon} {et~al.}(2010){Sharon}, {Gal-Yam}, {Maoz}, {Filippenko},
  {Foley}, {Silverman}, {Ebeling}, {Ma}, {Ofek}, {Kneib}, {Donahue}, {Ellis},
  {Freedman}, {Kirshner}, {Mulchaey}, {Sarajedini}, \& {Voit}}]{Sharon10}
{Sharon}, K., {et~al.} 2010, \apj, 718, 876

\bibitem[{{Silk} \& {Srednicki}(1984)}]{Silk84}
{Silk}, J., \& {Srednicki}, M. 1984, Physical Review Letters, 53, 624

\bibitem[{{Spergel} {et~al.}(2007){Spergel}, {Bean}, {Dor{\'e}}, {Nolta},
  {Bennett}, {Dunkley}, {Hinshaw}, {Jarosik}, {Komatsu}, {Page}, {Peiris},
  {Verde}, {Halpern}, {Hill}, {Kogut}, {Limon}, {Meyer}, {Odegard}, {Tucker},
  {Weiland}, {Wollack}, \& {Wright}}]{wmap07}
{Spergel}, D.~N., {et~al.} 2007, \apjs, 170, 377

\bibitem[{{Sreekumar} {et~al.}(1998){Sreekumar}, {Bertsch}, {Dingus},
  {Esposito}, {Fichtel}, {Hartman}, {Hunter}, {Kanbach}, {Kniffen}, {Lin},
  {Mayer-Hasselwander}, {Michelson}, {von Montigny}, {Muecke}, {Mukherjee},
  {Nolan}, {Pohl}, {Reimer}, {Schneid}, {Stacy}, {Stecker}, {Thompson}, \&
  {Willis}}]{Sreekumar98}
{Sreekumar}, P., {et~al.} 1998, \apj, 494, 523

\bibitem[{{Stecker}(1971)}]{Stecker71}
{Stecker}, F.~W. 1971, NASA Special Publication, 249

\bibitem[{{Stecker}(2007)}]{Stecker07}
---. 2007, Journal of Physics Conference Series, 60, 215

\bibitem[{{Stecker} \& {Jones}(1977)}]{Stecker77}
{Stecker}, F.~W., \& {Jones}, F.~C. 1977, \apj, 217, 843

\bibitem[{Stecker {et~al.}(1971)Stecker, Morgan, \&
  Bredekamp}]{Stecker71_matter}
Stecker, F.~W., Morgan, D.~L., \& Bredekamp, J. 1971, Phys. Rev. Lett., 27,
  1469

\bibitem[{{Stecker} {et~al.}(1993){Stecker}, {Salamon}, \&
  {Malkan}}]{Stecker93}
{Stecker}, F.~W., {Salamon}, M.~H., \& {Malkan}, M.~A. 1993, \apjl, 410, L71

\bibitem[{{Stecker} \& {Venters}(2011)}]{Stecker10}
{Stecker}, F.~W., \& {Venters}, T.~M. 2011, \apj, 736, 40

\bibitem[{{Strolger} {et~al.}(2004){Strolger}, {Riess}, {Dahlen}, {Livio},
  {Panagia}, {Challis}, {Tonry}, {Filippenko}, {Chornock}, {Ferguson},
  {Koekemoer}, {Mobasher}, {Dickinson}, {Giavalisco}, {Casertano}, {Hook},
  {Blondin}, {Leibundgut}, {Nonino}, {Rosati}, {Spinrad}, {Steidel}, {Stern},
  {Garnavich}, {Matheson}, {Grogin}, {Hornschemeier}, {Kretchmer}, {Laidler},
  {Lee}, {Lucas}, {de Mello}, {Moustakas}, {Ravindranath}, {Richardson}, \&
  {Taylor}}]{Strolger04}
{Strolger}, L.-G., {et~al.} 2004, \apj, 613, 200

\bibitem[{{Strong} {et~al.}(2000){Strong}, {Moskalenko}, \&
  {Reimer}}]{Strong00}
{Strong}, A.~W., {Moskalenko}, I.~V., \& {Reimer}, O. 2000, \apj, 537, 763

\bibitem[{{Strong} {et~al.}(2004){Strong}, {Moskalenko}, \&
  {Reimer}}]{Strong04}
---. 2004, \apj, 613, 962

\bibitem[{{Sullivan} {et~al.}(2006){Sullivan}, {Le Borgne}, {Pritchet},
  {Hodsman}, {Neill}, {Howell}, {Carlberg}, {Astier}, {Aubourg}, {Balam},
  {Basa}, {Conley}, {Fabbro}, {Fouchez}, {Guy}, {Hook}, {Pain},
  {Palanque-Delabrouille}, {Perrett}, {Regnault}, {Rich}, {Taillet}, {Baumont},
  {Bronder}, {Ellis}, {Filiol}, {Lusset}, {Perlmutter}, {Ripoche}, \&
  {Tao}}]{Sullivan06}
{Sullivan}, M., {et~al.} 2006, \apj, 648, 868

\bibitem[{{Tanaka} {et~al.}(2011){Tanaka}, {Allafort}, {Ballet}, {Funk},
  {Giordano}, {Hewitt}, {Lemoine-Goumard}, {Tajima}, {Tibolla}, \&
  {Uchiyama}}]{Tanaka11}
{Tanaka}, T., {et~al.} 2011, \apjl, 740, L51

\bibitem[{{Tang} \& {Wang}(2005)}]{Tang05}
{Tang}, S., \& {Wang}, Q.~D. 2005, \apj, 628, 205

\bibitem[{{Thompson} {et~al.}(2007){Thompson}, {Quataert}, \&
  {Waxman}}]{Thompson07}
{Thompson}, T.~A., {Quataert}, E., \& {Waxman}, E. 2007, \apj, 654, 219

\bibitem[{{Venters}(2010)}]{Venters10}
{Venters}, T.~M. 2010, \apj, 710, 1530

\bibitem[{{Venters} \& {Pavlidou}(2011)}]{Venters11}
{Venters}, T.~M., \& {Pavlidou}, V. 2011, \apj, 737, 80

\bibitem[{{Webbink}(1984)}]{Webbink84}
{Webbink}, R.~F. 1984, \apj, 277, 355

\bibitem[{{Weekes} {et~al.}(1989){Weekes}, {Cawley}, {Fegan}, {Gibbs},
  {Hillas}, {Kowk}, {Lamb}, {Lewis}, {Macomb}, {Porter}, {Reynolds}, \&
  {Vacanti}}]{Weekes89}
{Weekes}, T.~C., {et~al.} 1989, \apj, 342, 379

\end{thebibliography}

\end{document}